\documentclass[aps,showpacs,preprint]{revtex4}
\usepackage{epsfig}
\usepackage{dcolumn}
\usepackage{bm}

\normalsize

\begin{document}

\title{Quantized Topological Charges of  Ferroelectric Skyrmions  in  Two-Dimensional Multiferroic Materials}

\author{Zhaosen Liu\footnote{Email: liuzhsnj@yahoo.com, zsliu@hynu.edu.cn}}
\affiliation{
 College of Physics and Electronic Engineering, Hengyang
Normal University, Hengyang  421002, China
 }

\vspace{0.2cm}

\begin{abstract}
 In multiferroic materials, the  microscopic magnetic and electric textures are   strongly correlated with each other by  magnetoelectric (ME) coupling. Therefore, topological electric dipole textures, such as  ferroelectric (FE) dipole skyrmions and  periodic FE  dipole crystals,   are expected to be induced   if   ferromagnetic (FM) skyrmions and  FM   skyrmionic crystals  (or lattices so as  to be abbreviated as SLs for convenience) can be  stabilized.     In the present  work,     a quantum computational approach is utilized  to simulate the  topological textures of  FE   SLs in a  two-dimensional  (2D)  multiferroic  system.  Consequently, we find  that,   FE SLs can indeed be  induced  once the FM SLs are  formed;  each  FE  skyrmion is an ferroelectric dipolar complex   formed around an FM skyrmion;     the topological charges  of these  FE skyrmions are usually quantized to be integers, half integers and  multiples of  certain fractional values, so that  the  FE SL coincides  precisely with  the corresponding  FM SL;  the topological charge density   of each FE SL also forms periodic pattern;   and  a normally applied electric field is able to change the sizes of  FM skyrmions,   elevate their formation temperatures, and destroy  FM SLs below the critical temperatures  as well.
 \end{abstract}


\maketitle


\section{Introduction}

Magnetic skyrmions   were theoretically predicted by Bogdanov  $et.~al$ as stable magnetic states in  chiral magnets \cite{Bogdanov89, Bogdanov94}, and  discovered  in  helimagnetic conductor MnSi   about twenty years later \cite{Binz,Muhlbauer}.
These   quasi-particles   textures  built of topologically protected vortices \cite{4Nagaosa,12Fert13} are  mainly  caused by  Dzyaloshinskii-Moriya interactions (DMI)  arising from inversion symmetry breaking \cite{13Dzyaloshinskii,14Moriya,15Fert80,16Fert90,17Crepieux}.
Due to    their small sizes, topological  stability,  and  greatly reduced   electronic currents   required to be driven  to motion, magnetic skyrmions are believed  to be  promising  candidates  in future data storage  and  racetrack memory, thus have been  intensively studied    in recent years \cite{Pappas,Yu10,Banerjee,Yi,Buhrandt,  Huang,Romming15, Iwasaki}.

Magnetic skyrmions in metals can be created, deleted and driven to motion by applied spin electric currents based on the spin transfer torque (STT) mechanism
\cite{Jonietz,Schultz,Everschor,Zang}. However, the Ohmic heating incurred by the currents are undesirable for electronics, and the  STT technique does not work in insulators since  electric currents cannot   pass through.  Fortunately, the magneto-electric coupling  that is present  in   multiferroic insulators, such as Cu$_2$OSeO$_3$ \cite{Seki12},  opens  a new way to   manipulate   magnetic skyrmions   by means of  electric fields. In these materials,   ME coupling   arises  from the so-called {\it p-d} hybridization mechanism \cite{Seki12, LiuYH,LiuYH2,JiaCL}, and the   electric dipole moments  can be expressed in terms of magnetic  spins,  $ {\bf P} = \alpha\left(S_i^yS_i^z, S_i^zS_i^x, S_i^xS_i^y\right)$. Consequently,   the   topological  FM and FE   chiral  textures are expected to be  present   simultaneously \cite{Seki12,Tokura}.

Based on  above theory,  the chiral  textures of electric dipoles and dipolar charge density in   Cu$_2$OSeO$_3$  were simulated,   by two research groups,    within  external magnetic  and  electric fields   applied in various  directions \cite{Seki12,LiuYH}. The  calculated  electric dipole textures   exhibited  regular and distinct topological  characters. Therefore,  it is of great interest  to further study their    topological properties.

For this  purpose,    the FE  SLs of a   2D   multiferroic  material are calculated  here  by means of   a quantum computational approach \cite{liujpcm,liupssb,  LiuIan19-2,LiuIan20-SM},    their topological properties are then  further analyzed and  characterised.  Consequently, we find   that an FE SL is induced once one FM SL is stabilized; each FE skymrion is   an electric dipole complex   formed around  an FM skyrmion,  and the topological charges  of these FE skyrmions are usually quantized to be integers, half integers and multiples of  certain fractional values;  the topological charge density   of  every FE SL also forms periodic pattern; and   normally applied electric fields are  able to change  the sizes of  FM skyrmions,   elevate the  formation temperatures of  FM SLs, and destroy  them below the critical temperatures as well.

\section{Theoretical Model and  Method}

The  2D multiferroic material, simplified here as a monolayer,  is  assumed   to be  in the $xy$-plane.  Its  Hamiltonian,  ${\cal H}  = {\cal H}_{M} +{\cal H}_{ME}$,   consists of  two parts,  which  are originated from various   magnetic  interactions and ME coupling respectively. The  first  part  can be   written as
{\small
\begin{eqnarray}
{\cal H_M} = &  -\frac{1}{2}\sum_{i,j}\left[{\cal J}_{ij}{ \bf{S}_i
\cdot }\bf{S}_j
+ {\bf{D}_{ij}\cdot(\bf{S}_i\times}\bf{S}_j)\right]
    - {\bf H }\cdot\sum_i{\bf S}_i
-K_A\sum_i\left(\bf{S}_i\cdot \hat{n}  \right)^2\;,
\label{hamil}
\end{eqnarray}
}
where   ${\cal J}_{ij}$,  $ \bf {D}_{ij}$,  $K_A$ are the  strengths of  HE, DM  and  uniaxial anisotropic   (UMA) interactions  respectively. While the   portion resulting from   ME coupling  of strength $\alpha$   can be expressed with
 \begin{eqnarray}
 {\cal H}_{ME}  = -\alpha\sum_i{\bf P_i}\cdot{\bf E_i}= -\frac{\alpha}{2}\sum {\bf S}_i\left(
 \begin{array}{ccc}
 0          & E_i^z & E_i^y\\
 E_i^z  & 0 &  E_i^x\\
 E_i^y & E_i^x & 0\\
 \end{array}
 \right){\bf S}_i \;.
 \end{eqnarray}

Since quantum theory is employed,  spins  appearing in  above  Hamiltonian  are  quantum operators \cite{liujpcm,liupssb,LiuIan19-2,LiuIan20-SM};   the thermal average of every  physical observable $A$  at temperature $T$  is   evaluated   with
\begin{equation}
\langle A\rangle = \frac{{\rm Tr}\left[\hat{A}\exp(\beta{\cal H}_i)\right]}{{\rm Tr}\left[\exp(\beta{\cal H}_i)\right]}\;,
\label{avq}
\end{equation}
where   $\hat{A}$ is the  operator of  observable $A$,   $\beta = -1/k_BT$;  and    $[S_{\alpha},S_{\beta}] = i{\hbar}S_{\gamma}\epsilon_ {\alpha,\beta, \gamma}$ with $\alpha,\beta, \gamma = x,y,z$,    $\epsilon_ {\alpha,\beta, \gamma}$ = 0, $\pm 1$,  depending on the order of  $ \alpha$, $\beta$ and $ \gamma$. In contrast,   two arbitrary  spin components simply commute in classical physics.  Therefore,  the  two sorts of   methods  are different from each other. Especially, the uncertainty principle of quantum theory   enables  quantum computing codes to tunnel through  energy barriers much easily, so as to converge to the equilibrium states  without the need to   manipulate  the spins elaborately.

 \section{Computational Results}

We have previously simulated the chiral spin textures  of a  2D FeSi$_{0.5}$Co$_{0.5}$Si-like FM thin film \cite{LiuIan20-SM}. With    $D/{\cal J}$ assigned to 1.027 and external   magnetic  fields  of strengths $H \in$ (0.15,0.37)  applied  perpendicularly,  FM SLs, as shown in  Ref.\cite{Yu10},   could be reproduced at  low temperatures \cite{LiuIan20-SM}.  In the present work,  both ${\bf H}$ and ${\bf E}$ are considered  to be applied normally  to the 2D multiferroic monolayer of the square structure,  where every lattice site  is occupied by an  $S$ = 1 spin,    $\alpha$ is set to 0.2,    while  $D/{\cal J}$   to  the above same value \cite{LiuIan20-SM,Yu10}    for direct  comparison. This  $D/{\cal J}$  ratio is  very close to   that   used by Liu {\it et al.}  in their Monte Carlo simulations for another  2D multiferroic system \cite{LiuYH2} .

 \subsection{Effects of Electric Fields on   Sizes and Formation Temperatures of Magnetic Skyrmions}
\vspace{-0.3cm}

Indeed, external electric fields are able to manipulate magnetic skyrmions through the ME coupling as shown  in Figure 1, where the skyrmion sizes are changed and formation  temperatures   of FM SLs elevated by  the electric fields that are  applied perpendicularly to the $xy-$plane.

In Figure 1(a),   only an external magnetic field with strength $H$  = 0.2 is exerted, FM SLs are  formed below critical temperature $T_{SL} $ = 3.15, and each skyrmion in these FM SLs occupies $N_P$ =  50 lattice sites in average.   While    an electric field is also   exerted,  $T_{SL} $ is  firstly elevated  to  3.16 when $E \in$ (0.2,0.5), then to 3.17   as $E \in$ (0.6, 0.9). Meanwhile, $N_P$ is increased from 50 to 60, then to 64.8  in the two $E$-field ranges respectively. However, when   $E$  is increased to  1.0,  FM SL can only survive in a narrow high temperature interval $T \in$ (3.10, 3.17);   below this interval,  FM SL is replaced by FM helical crystal (HL)  which  persists down to very low temperatures. If  $E$  is further increased   to 1.2,    only FM HL can be observed     at low temperatures.

\begin{figure*}[htb]
\centerline{
 \epsfig{file=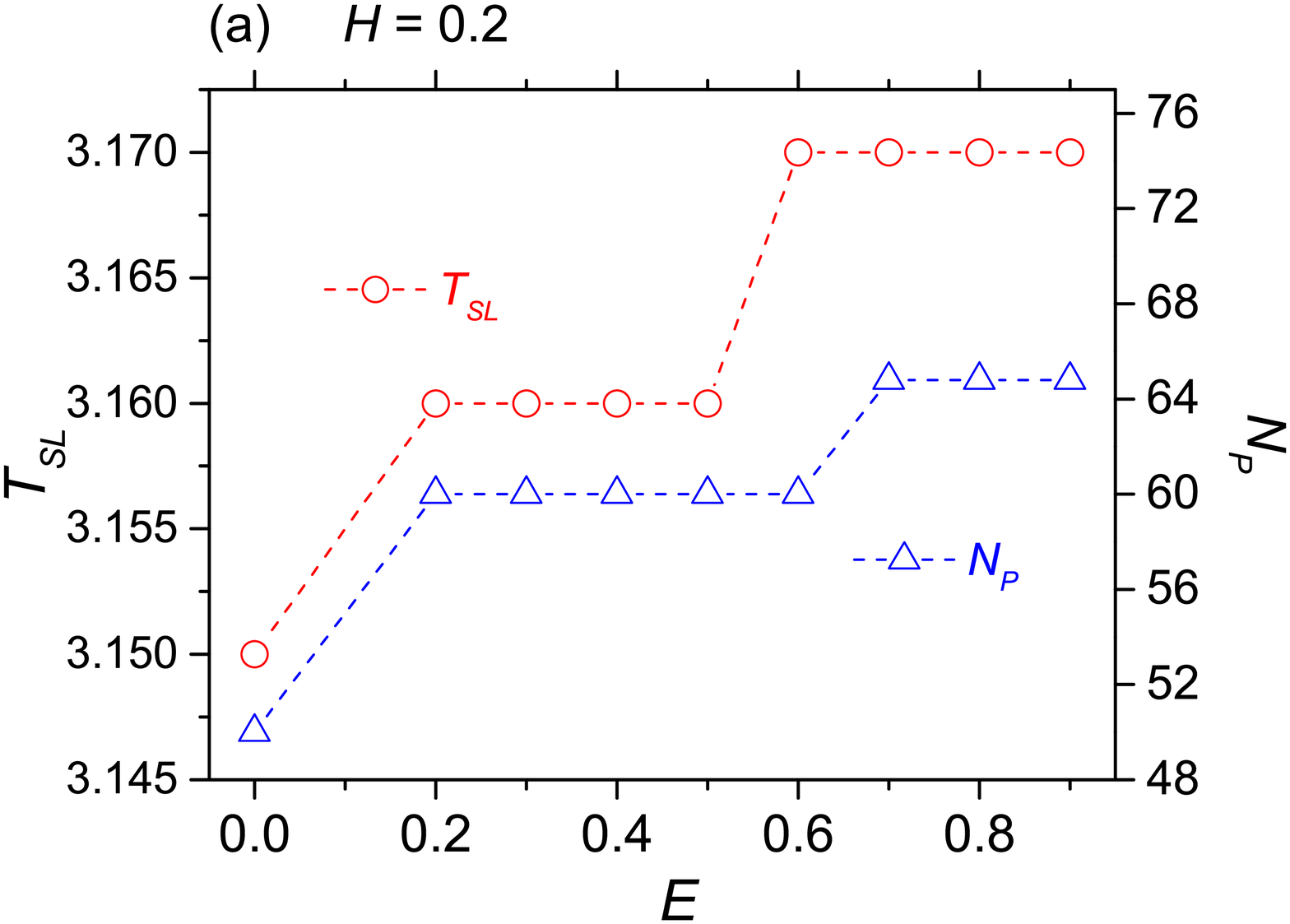,width=0.25\textwidth,height=4.5cm,clip=}
 \epsfig{file=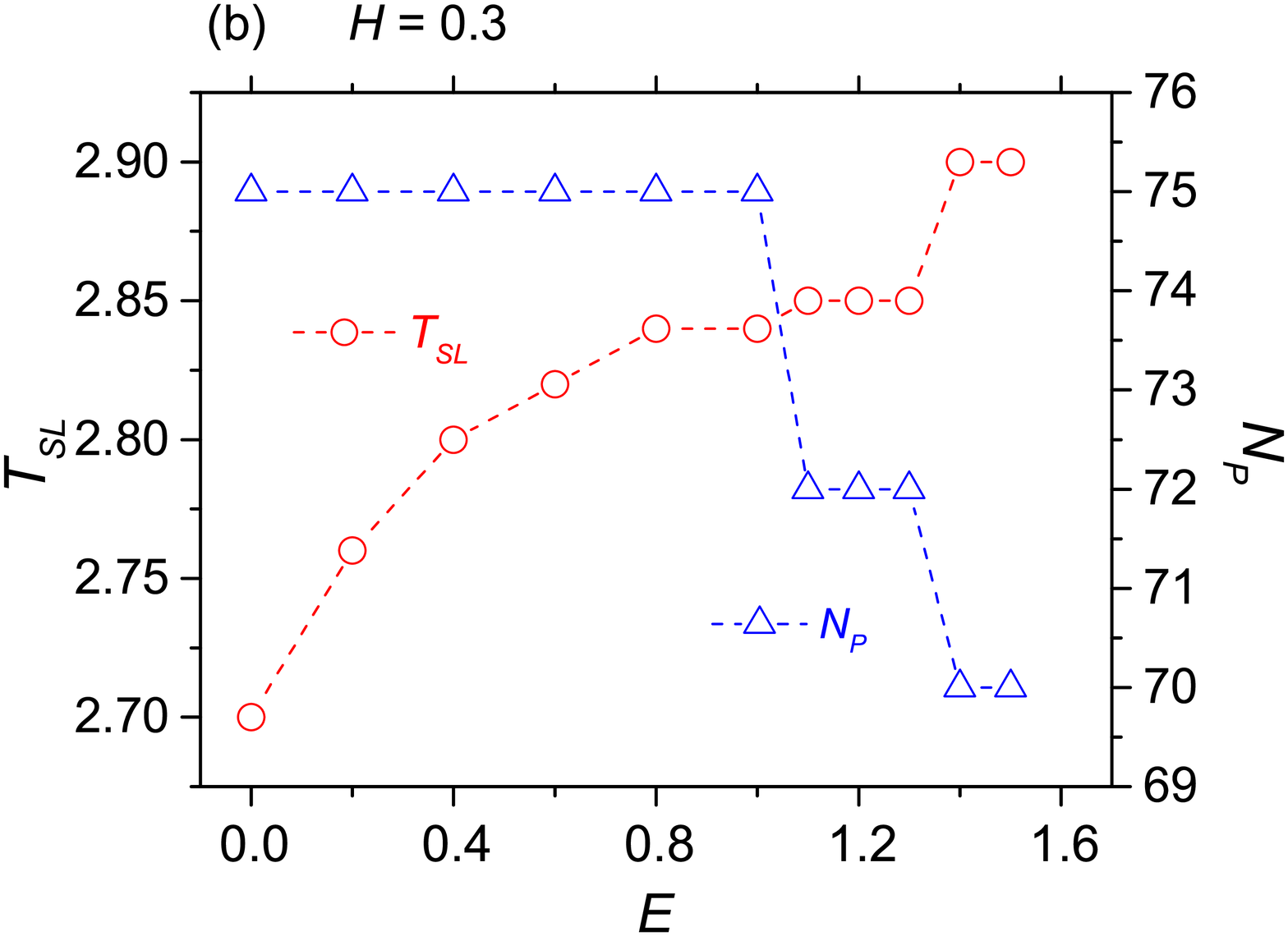,width=0.25\textwidth,height=4.5cm,clip=}
 \epsfig{file=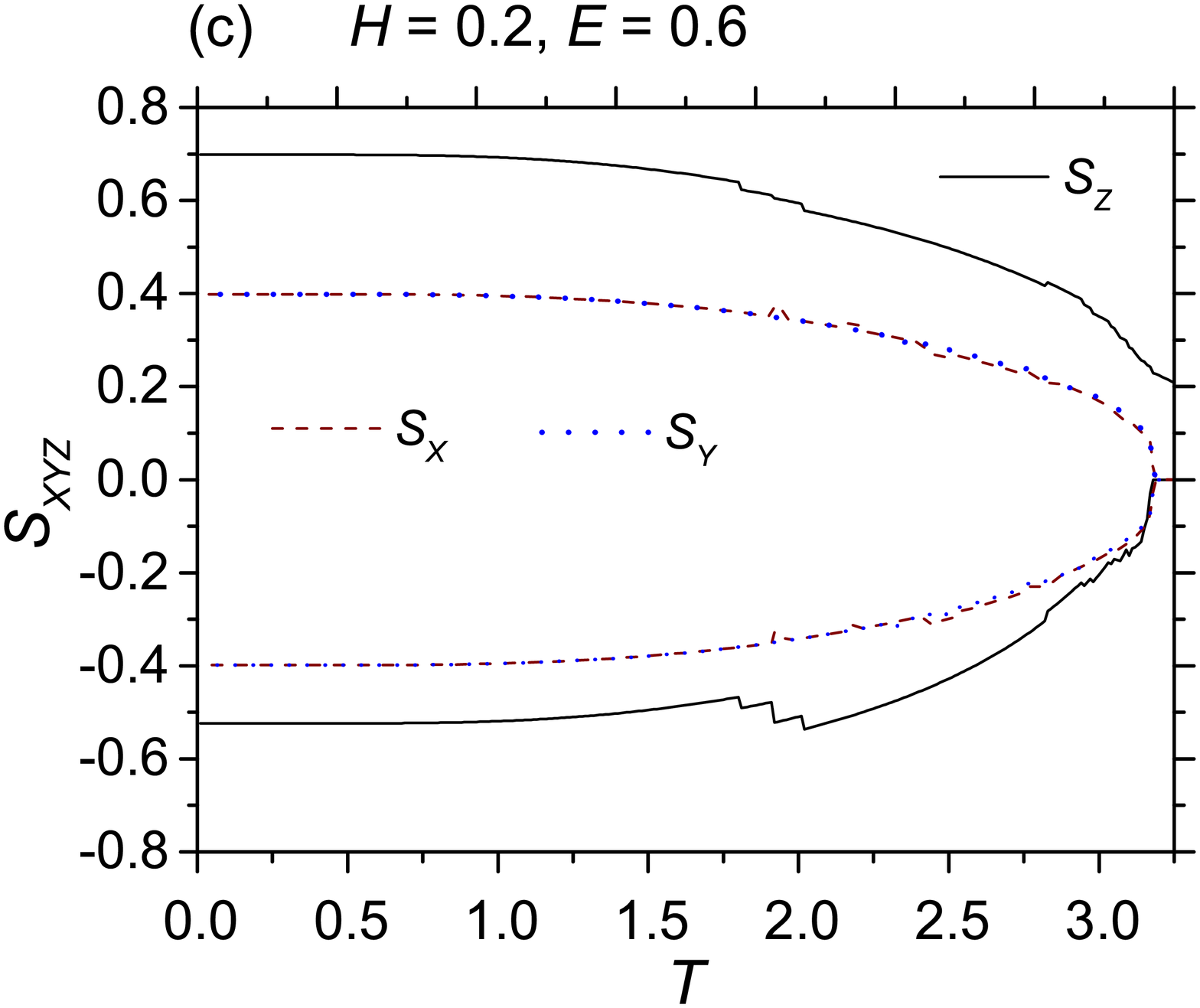,width=0.25\textwidth,height=4.5cm,clip=}
 \epsfig{file=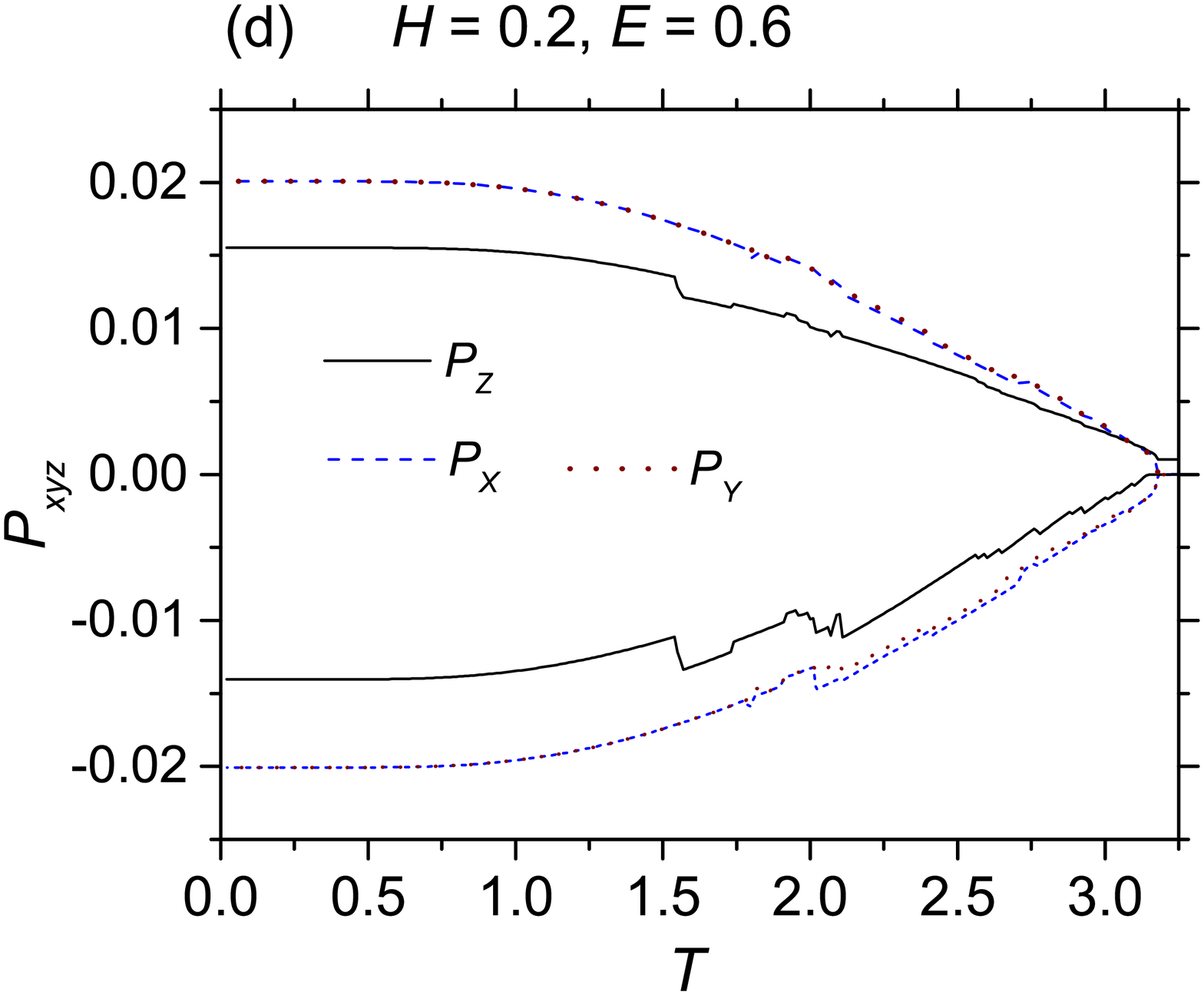,width=0.25\textwidth,height=4.5cm,clip=}
 }
  \vspace{-0.5cm}
\begin{center}
\parbox{16.5cm}{\small{{\bf Figure 1.}   Formation temperatures of FM SLs and spins occupied by an FM skyrmion versus normally  applied electric field $E$ when  (a,b) $H$ = 0.2, and 0.3 respectively;  Thermally averaged values of  three components of  (c) magnetic spins, and (d) electric polarizations   as functions of temperature while $H$ = 0.2 and $E$ = 0.6, respectively.}}
\end{center}
\end{figure*}

 \vspace{-0.3cm}
In contrast, when $H$ = 0.3, FM SLs can be stabilized in a wider range $E \in$ (0, 1.6)  as shown in Figure 1(b);  and  as $E$ increases,     $T_{SL}$   is elevated considerably  from 2.70 to 2.90,  but  $N_P$  decreases  from 75 to 72 and then to 70 unexpectedly, that is,   skyrmion size shrinks in this  process.

We have seen that  an applied $E$-field   can  strongly affect the microscopic structures of magnetic  skyrmions.   However,  it   has much less  effects upon the macroscopic magnetic properties. In Figure 1(c),  $E$ = 0.6,  the thermally averaged values of $\langle S_X \rangle$, $\langle S_Y \rangle$ and $\langle S_Z \rangle$ are plotted. In comparison with those curves obtained free of   $E$-field  interaction which are not presented,  the  three curves   displayed here have  actually     been slightly  modified, but   become  a little   smoother.   That is,  $E$-fields are able to suppress thermal disturbances  and   stabilize the FM and FE  microscopic structures.

 As expected,  once an FM SL is induced  below the formation temperature $T_{SL}$,  an  FE SL appears  immediately  due to the  ME coupling,  and the   microscopic configuration  of this FE SL can be changed by applied $E$-fields    for the same reason.  Figure 1(d) displays the thermal averaged values of $\langle P_X \rangle$, $\langle P_Y \rangle$ and $\langle P_Z \rangle$ versus changing temperature. The two sets of  curves that are calculated   with the same set of parameters  exhibit very similar characters, however  $|\langle P_Z \rangle|$ is  much  smaller than other two components because of  the extremely  tiny thickness of the monolayer. Similarly,    the   $P_{X,Y,Z}$ curves    have only been slightly enhanced   in magnitude by the $E$-field.

\begin{figure*}[htb]
\centerline{
 \epsfig{file=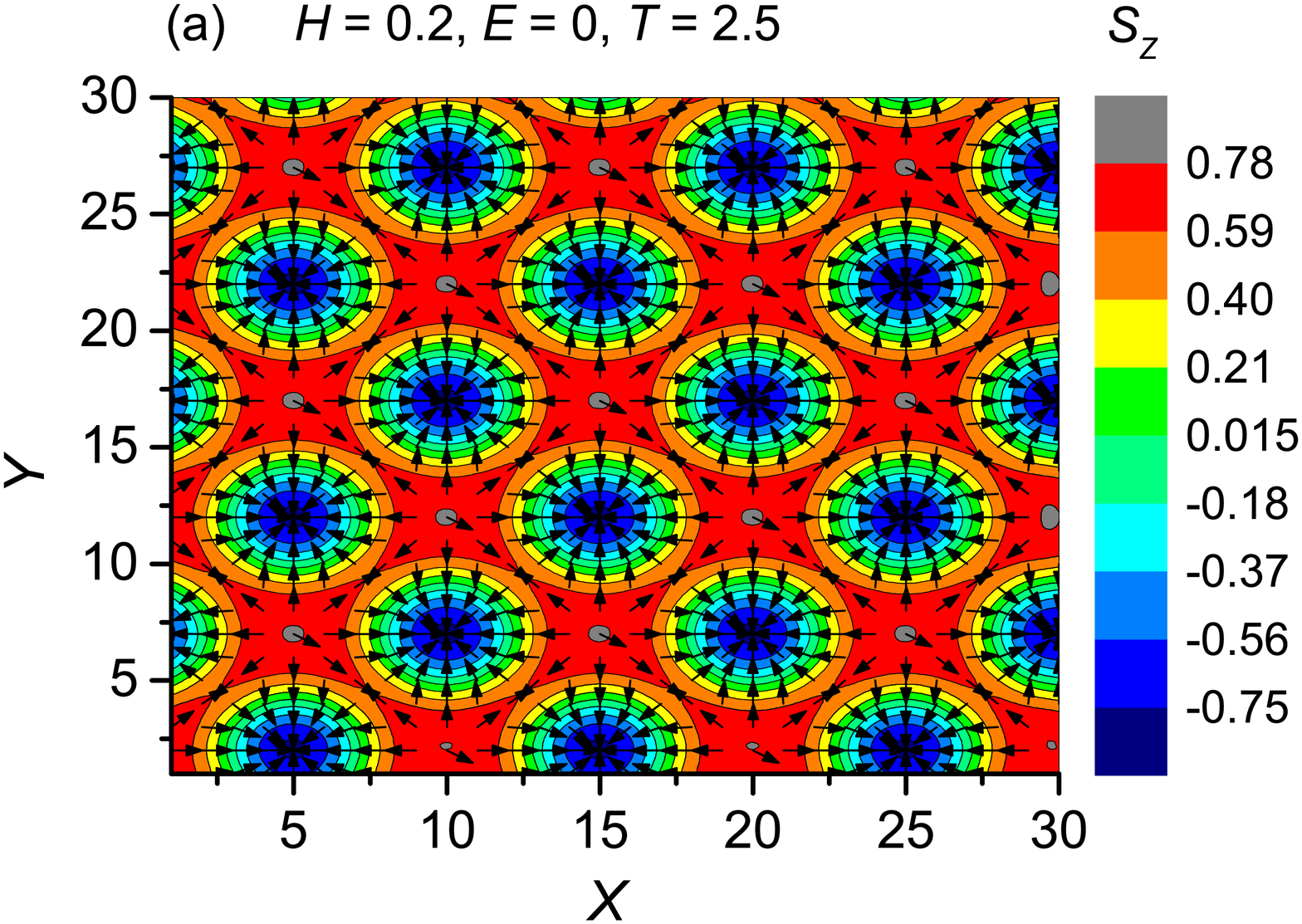,width=0.33\textwidth,height=5.2cm,clip=}
 \epsfig{file=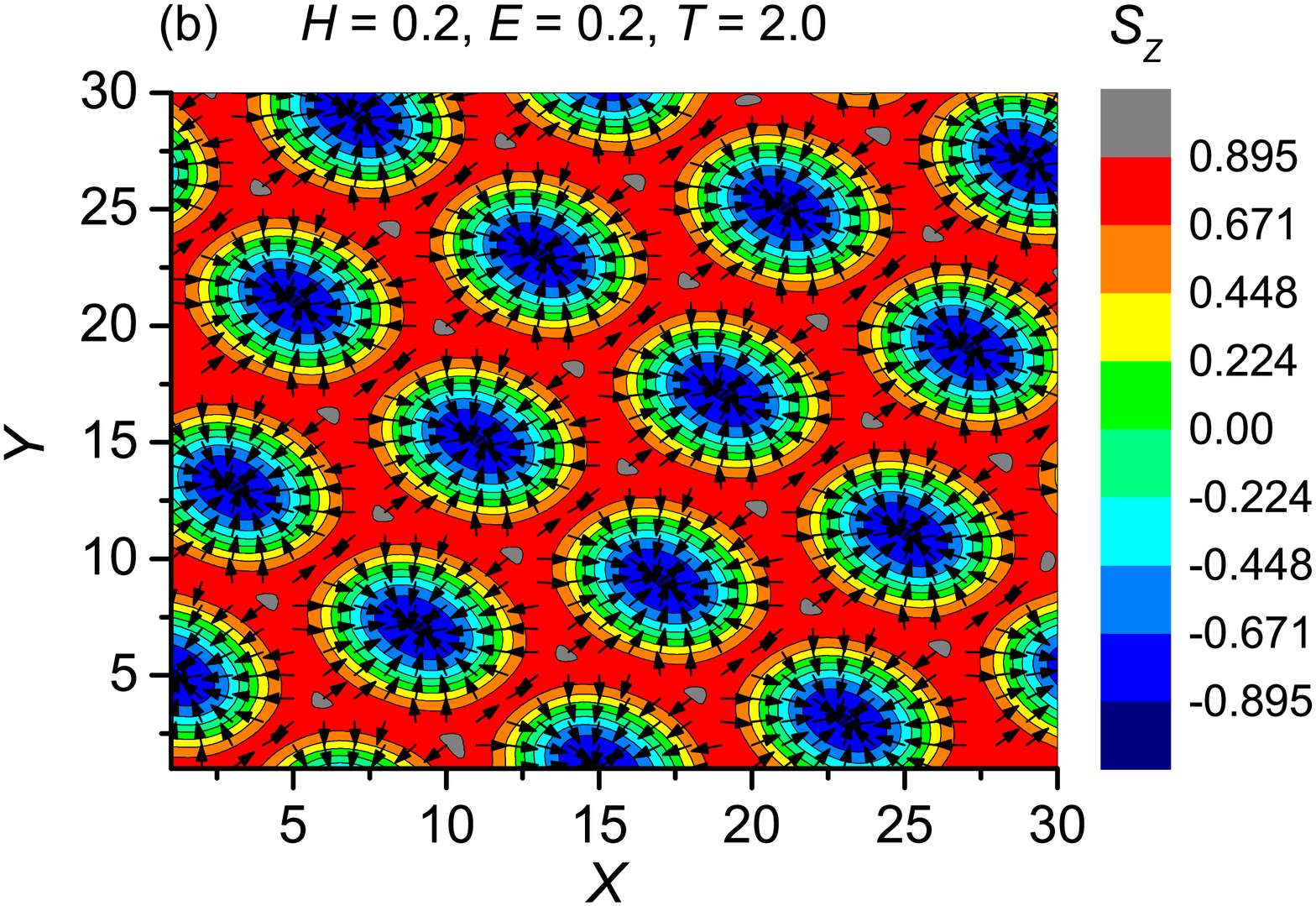,width=0.33\textwidth,height=5.2cm,clip=}
 \epsfig{file=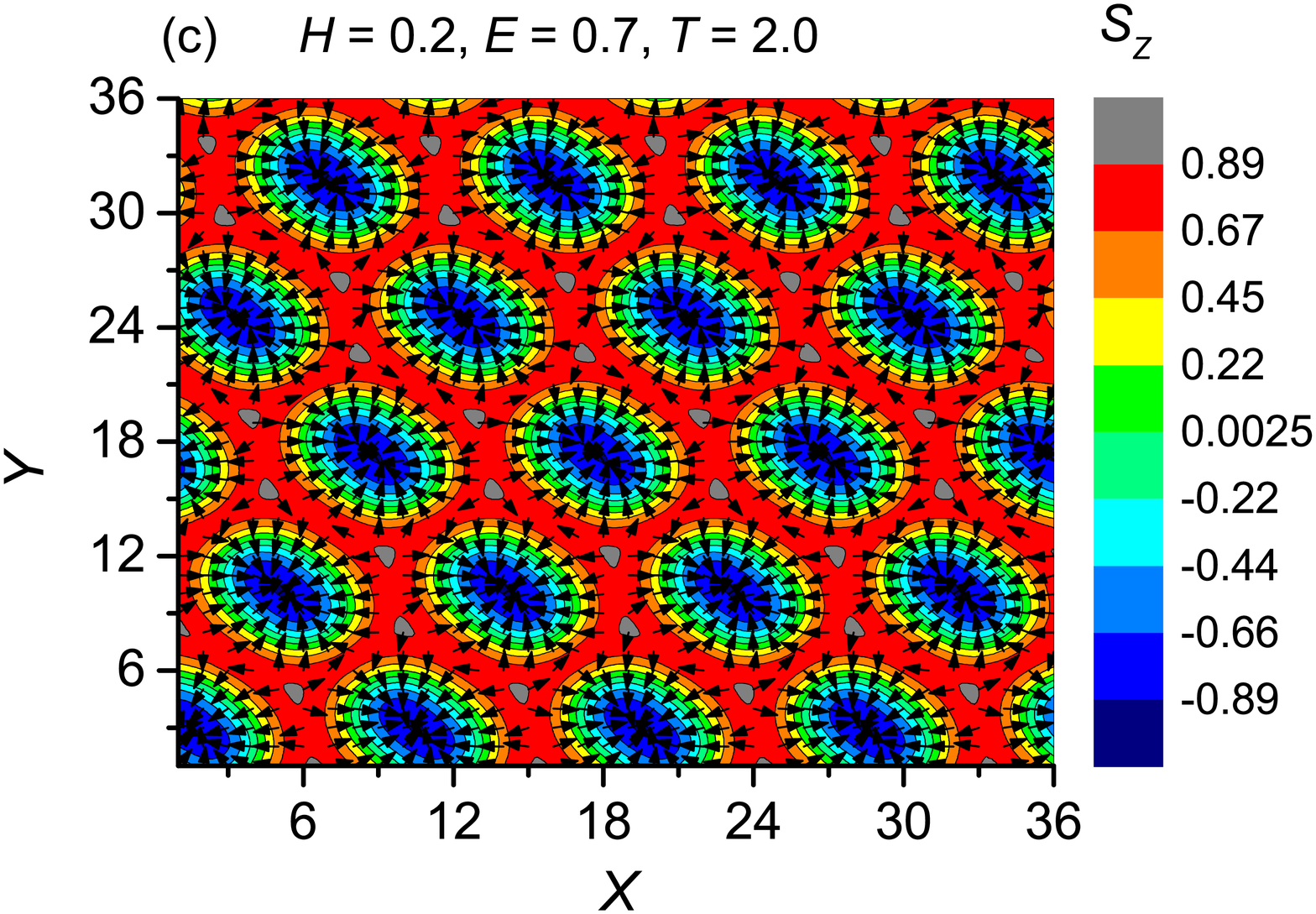,width=0.33\textwidth,height=5.2cm,clip=}
 }
 \vspace{-0.5cm}
\begin{center}
\parbox{16.5cm}{\small{{\bf Figure 2.}  FM SLs calculated with  $E$ equal to   (a) 0, (b) 0.2, and (c)   0.7 respectively  at different temperatures, while  $H$ = 0.2 in all of these cases.  }}
\end{center}
\end{figure*}

The  effects of   $E$-fields upon the microscopic structures of  FM SLs  are rather pronounced.  In Figure 2, $H$ is   fixed to 0.2,  while $E$-value differs.   This gives rise  to changed sizes of skyrmions   which   can be easily determined   from the numbers of skyrmions observed   and the  side-lengths  of the square lattices chosen for simulations.  On the other hand, when  no $E$-field is applied, 18 FM skyrmions condense to  an FM SL in  the triangular pattern; as $E$ is increased to  0.2,    the FM SL consisting of  15 elongated skyrmions  shows a rotated hexagonal-closely-packed (HCP) structure; while  $E$ = 0.7, the HCP  unit cell   is  twisted,  meanwhile   the 20  skyrmions are further elongated.

\subsection{Quantized  Topological Charges of Ferroelectric Skyrmions}

  In     continuous models,     every  magnetic skyrmion can be characterised by  an integer number,  which is  referred to as  topological  charge,  and can be expressed with 
\begin{equation}
Q_M = \frac{1}{4\pi}\int\int
{\bf  m}\cdot\left(\partial_x{\bf  m}\times\partial_y{\bf  m}\right)dxdy
\label{QN}\;,
\end{equation}
   and  whose   density is given by
  \begin{equation}
\rho = \frac{1}{4\pi}\bf {m}\cdot\left(\partial_x\bf {m} \times\partial_y\bf {m}\right)\;.
\label{rhoc}
\end{equation}

  However, our model is discrete and quantized,   the  topological charges of skyrmions  have to be calculated with another  set of formulas derived by Berg and his coworker  \cite{Berg}.   For   a unit square cell  with   four  spins $\bf { S}_1$, $\bf { S}_2$, $\bf { S}_3$,  and $\bf { S}_4$  at the corners, the topological charge density at the center  $x^*$   is given by
\begin{equation}
\rho(x^*) = \frac{1}{4\pi}\left[(\sigma A) (\bf {S}_1,\bf {S}_2,\bf {S}_3)+(\sigma A) (\bf {S}_1,\bf {S}_3,\bf {S}_4) \right]\;,
\label{rho}
\end{equation}
where $(\sigma A) (\bf {S}_i,\bf {S}_j,\bf {S}_k)$ stands for the signed area of the spherical triangle with corners $\bf {S}_i,\bf {S}_j,\bf {S}_k$,  and such a signed area $(\sigma A)(\bf {S}_1,\bf {S}_2,\bf {S}_3)$   is evaluated  with
\begin{eqnarray}
\exp[\frac{i}{2}(\sigma A)] = \gamma^{-1}\left[1+\bf {S}_1\cdot\bf {S}_2 +\bf {S}_2\cdot\bf {S}_3+\bf {S}_3\cdot\bf {S}_1
+i\bf {S}_1\cdot(\bf {S}_2\times\bf {S}_3)  \right]\;,\nonumber\\
\gamma = \left[ 2(1+ \bf {S}_1\cdot\bf {S}_2)(1+ \bf {S}_2\cdot\bf {S}_3)(1+ \bf {S}_3\cdot\bf {S}_1)\right]^{1/2} > 0\;.
\label{area}
\end{eqnarray}

\begin{figure*}[htb]
\centerline{
\epsfig{file=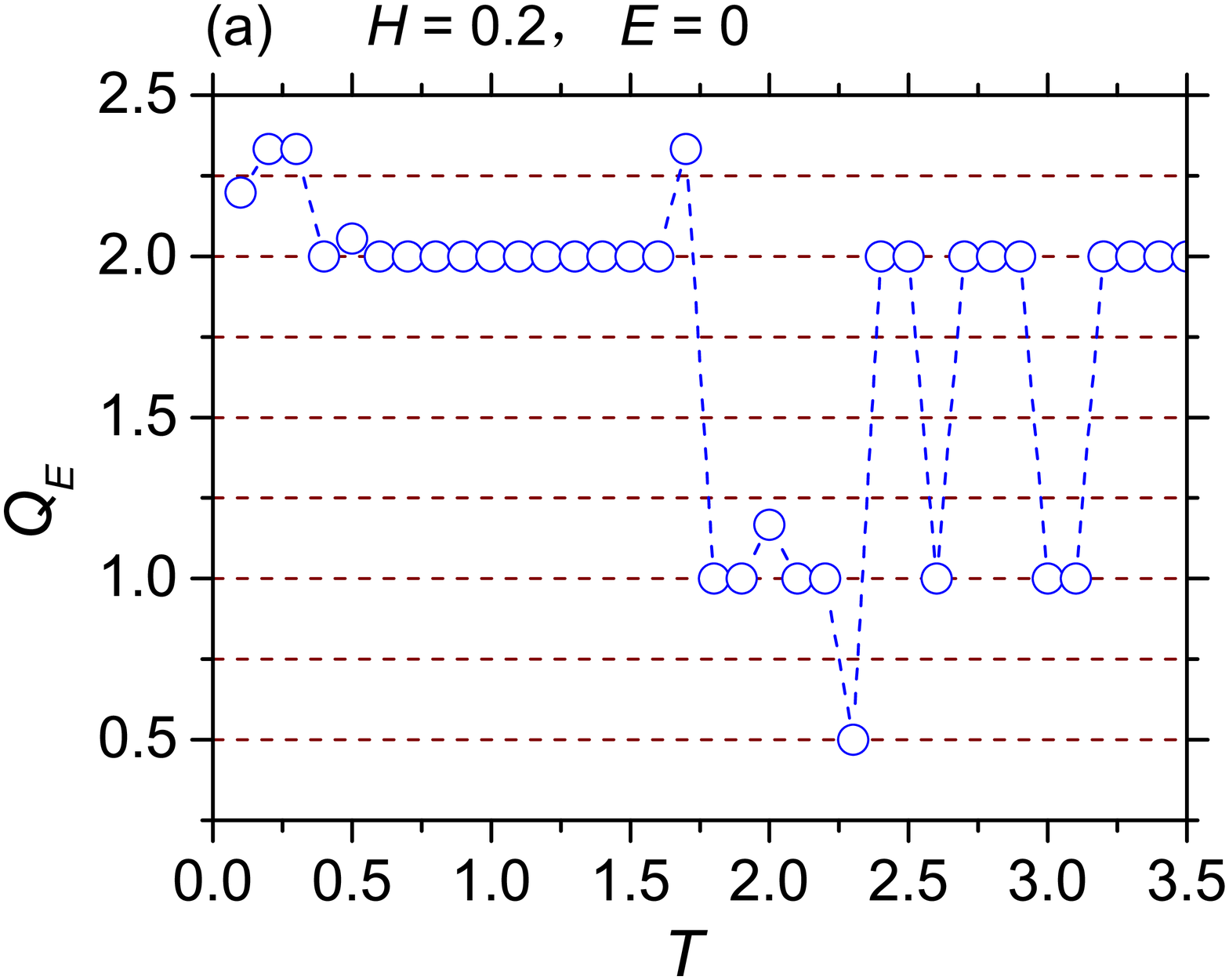,width=0.33\textwidth,height=4.5cm,clip=}
 \epsfig{file=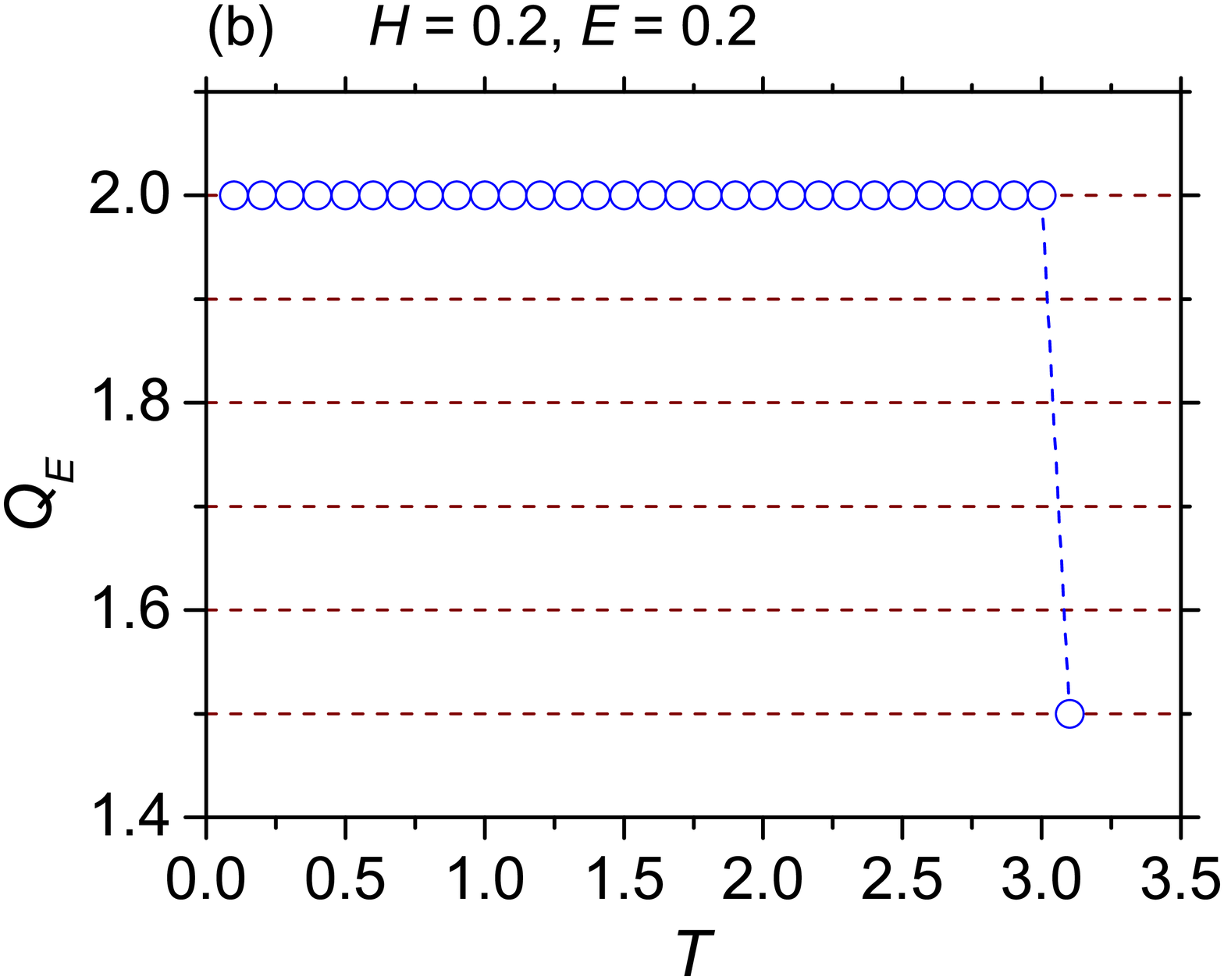,width=0.33\textwidth,height=4.5cm,clip=}
 \epsfig{file=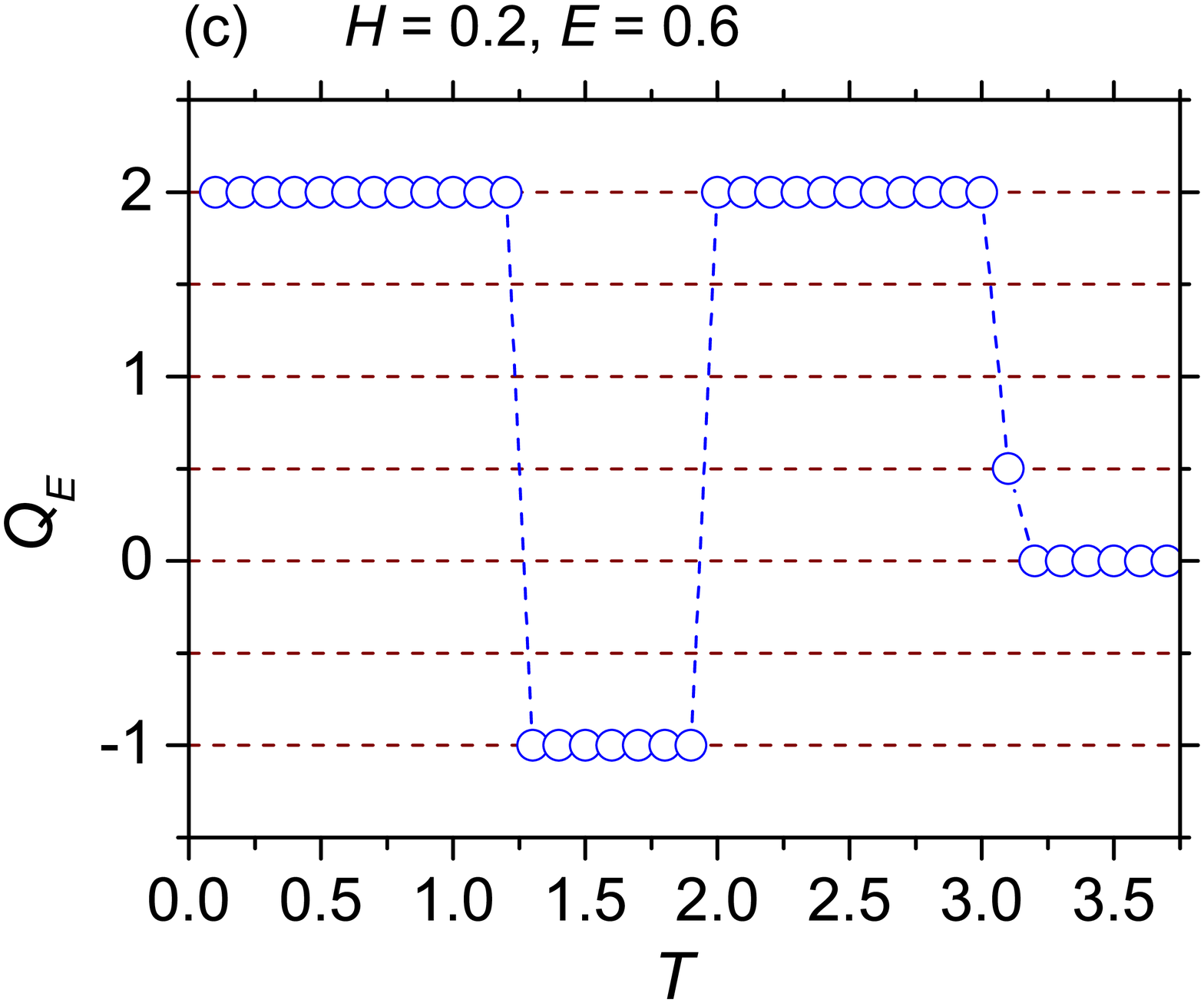,width=0.33\textwidth,height=4.5cm,clip=}
 }
 \centerline{
\epsfig{file=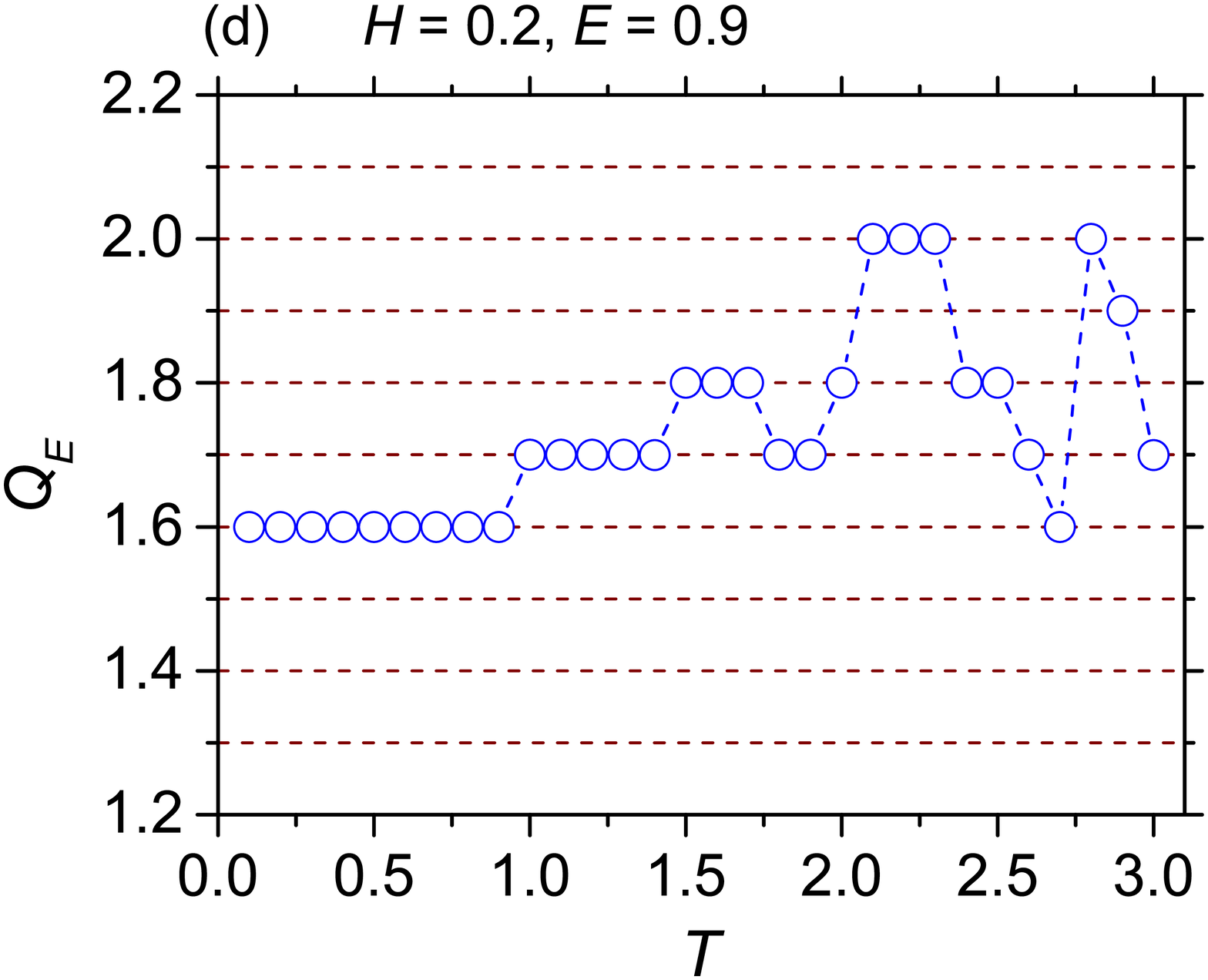,width=0.33\textwidth,height=4.5cm,clip=}
 \epsfig{file=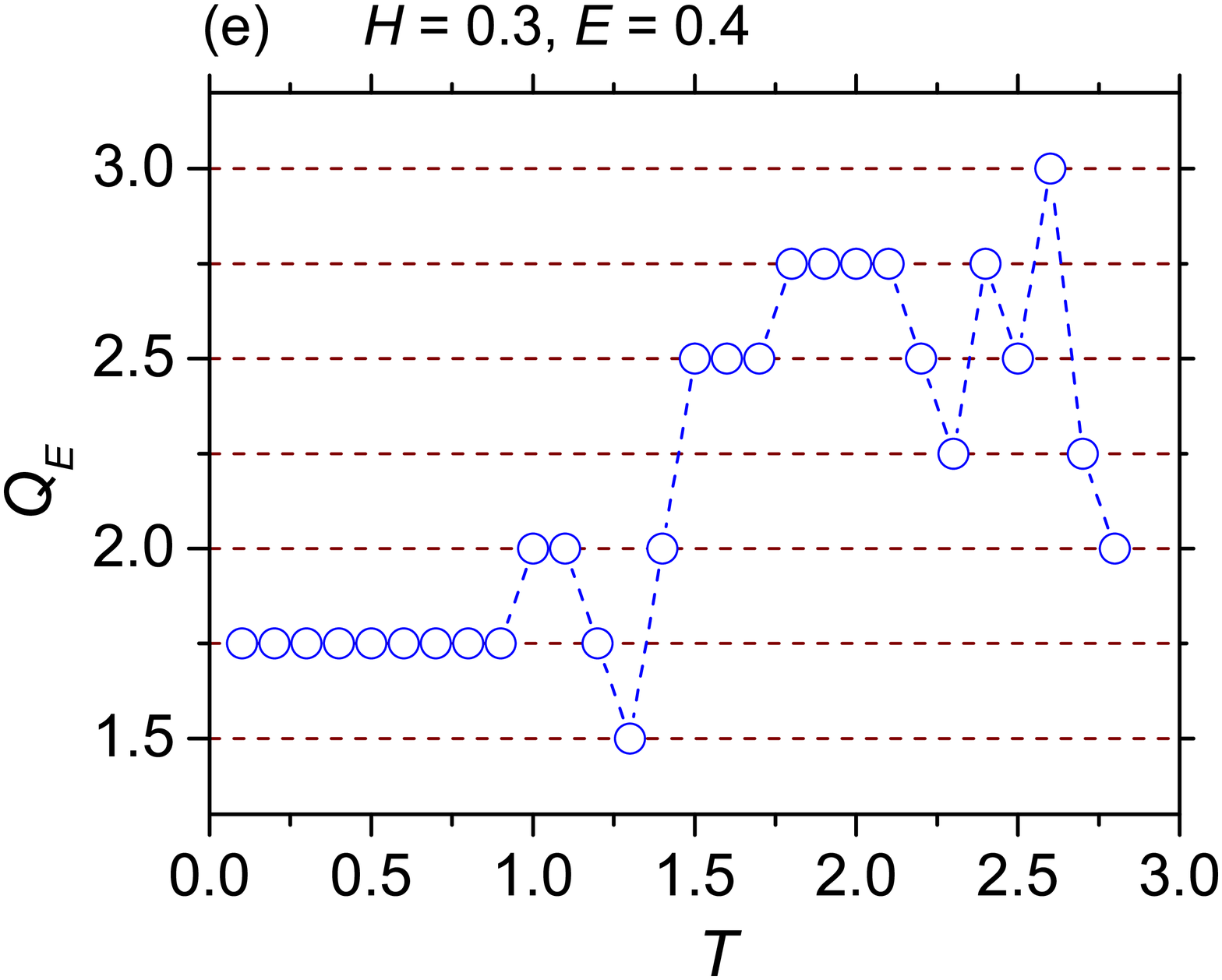,width=0.33\textwidth,height=4.5cm,clip=}
 \epsfig{file=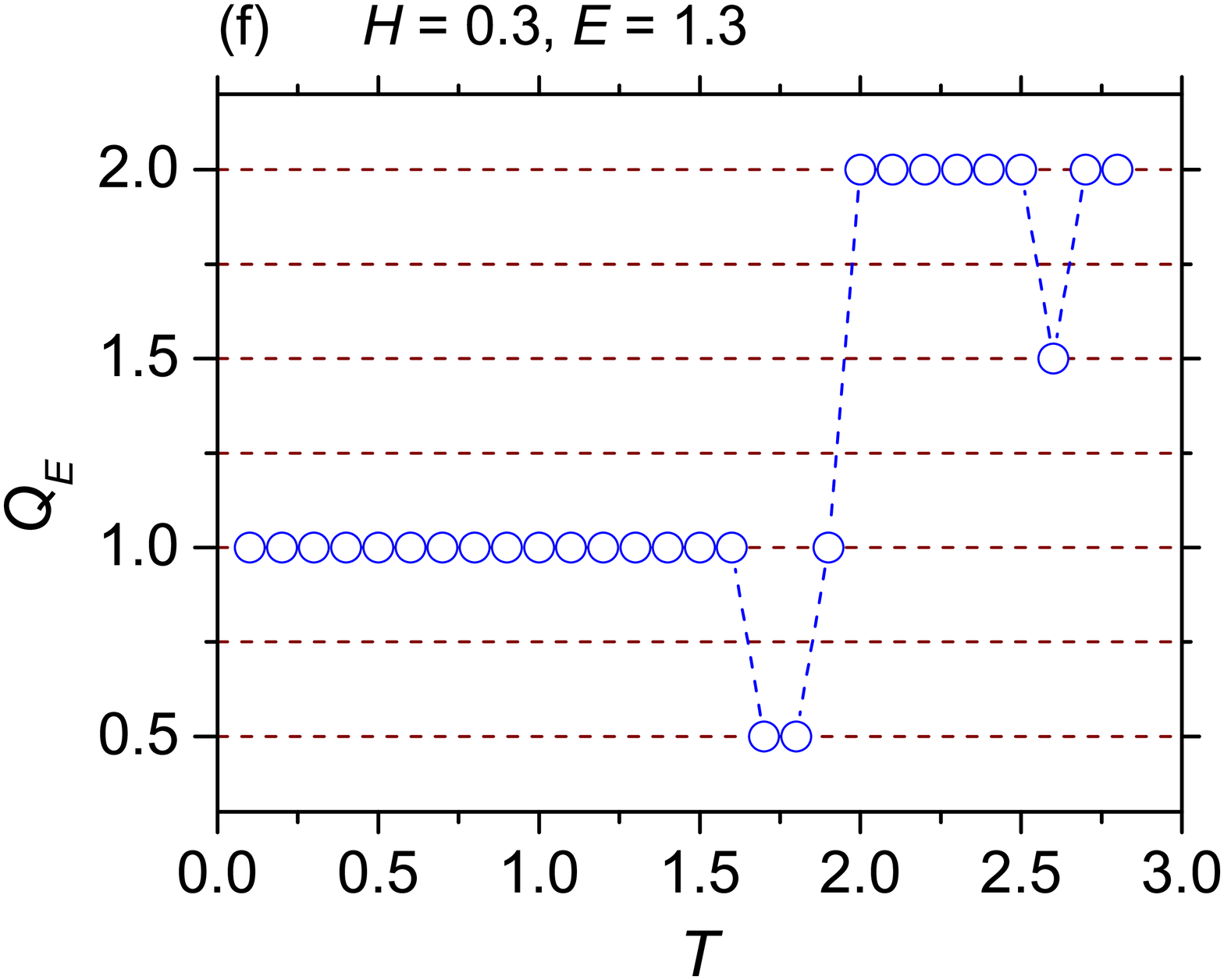,width=0.33\textwidth,height=4.5cm,clip=}
 }
 \vspace{-0.3cm}
\begin{center}
\parbox{16.5cm}{\small{{\bf Figure 3.}  Topological charge per skyrmion of the FE SL calculated with   (a) $H$ = 0.2,  $E$ = 0,  (b) $H$ = 0.2,  $E$ = 0.2,  (c) $H$ = 0.2,  $E$ = 0.6,  (d) $H$ = 0.2,  $E$ = 0.9,  (e) $H$ = 0.3,  $E$ = 0.4,  and  (f) $H$ = 0.3,  $E$ = 1.3, respectively, versus changing  temperature.}}
\end{center}
\end{figure*}

Using the  above formulas,  the   topological charge per FM skyrmion, $Q_M$, of every  FM SL is calculated to be -1.0, though  our model is quantized and  discrete. Obviously,  this set of  formulas can be employed to evaluate the topological charges of FE skyrmions as well. Interestingly, as depicted    in Figure 3,   most  obtained  $Q_{E}$ values  are integers (-1.0, 0, 2.0, 3.0), half integers  (0.5, 1.5, 2.5), and  odd multiples of 0.25 (1.75, 2.25, 2.75), which  form  horizontal steps in the curves. More  strangely,     the allowed  $Q_{E}$ values can also  be  1.6, 1.7, 1.8, 1.9, 2.0,  which are the multiples of 0.1, and only when $E$ = 0, $Q_{E}$ can take other  values.

\subsection{Ferroelectric Skyrmionic Crystals with $Z$-Contour}
\vspace{-0.3cm}
Figure 4 displays the   FE SLs  calculated with and without  $E$-fields applied. As coupled by  ME interaction,     FM SL and  FE SL appear   simultaneously, and  every FE skyrmion is formed around an FM skyrmion, so that a pair of   FE SL and FM SL contain  same number of FE skyrmions and  FM skyrmions respectively. Each    FE skyrmion   is an electric dipole complex consisting of, for example, two red areas (of  positive  $P_Z$ values) and two blue  regions (of  negative   $P_Z$ values), which is   a  typical feature of  the FE skyrmion as calculated   by Seki {\it et al.} though  there the  FM SLs are the Bloch-type    \cite{Seki12}.
\begin{figure*}[htb]
\centerline{
\epsfig{file=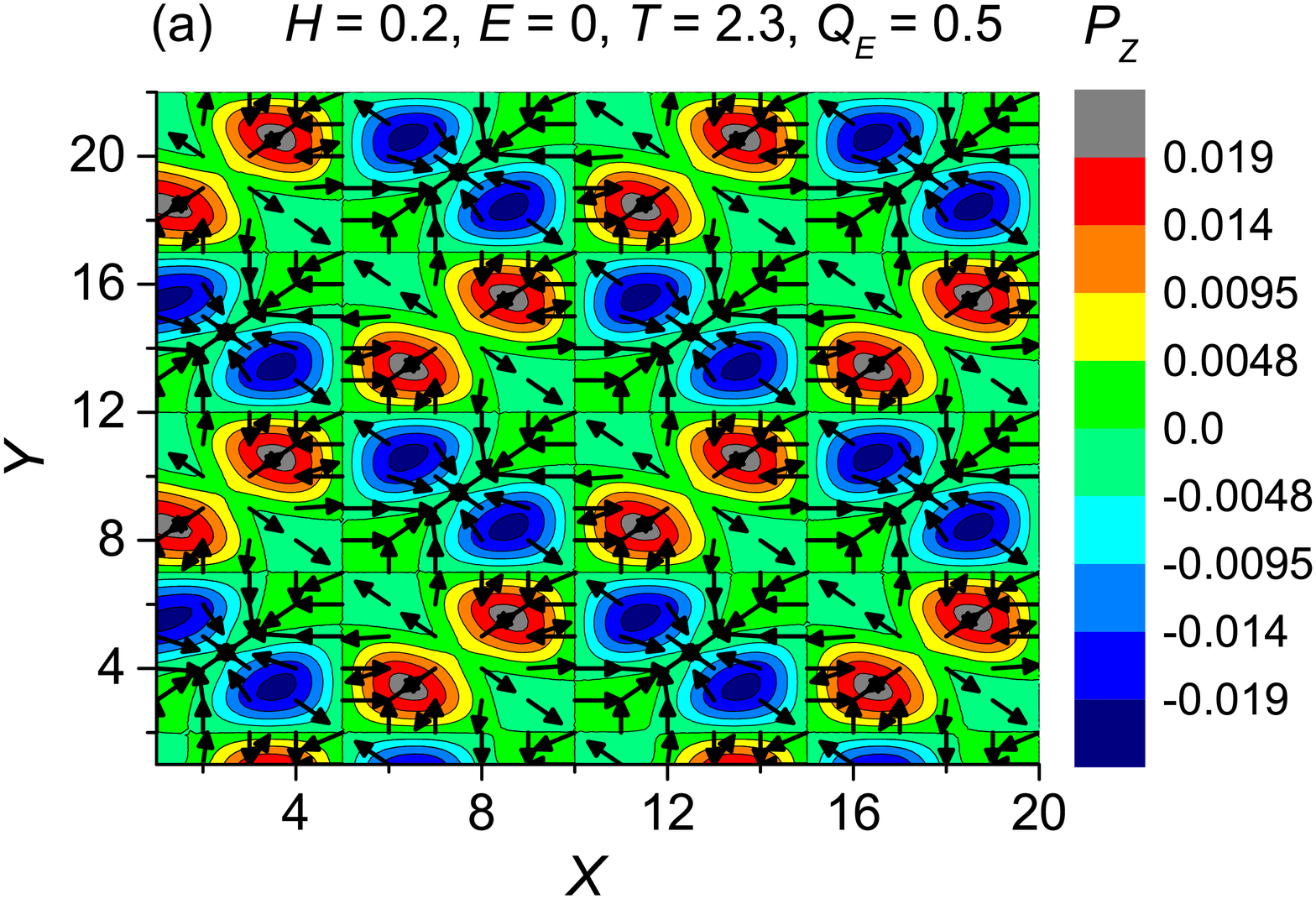,width=0.33\textwidth,height=5.2cm,clip=}
 \epsfig{file=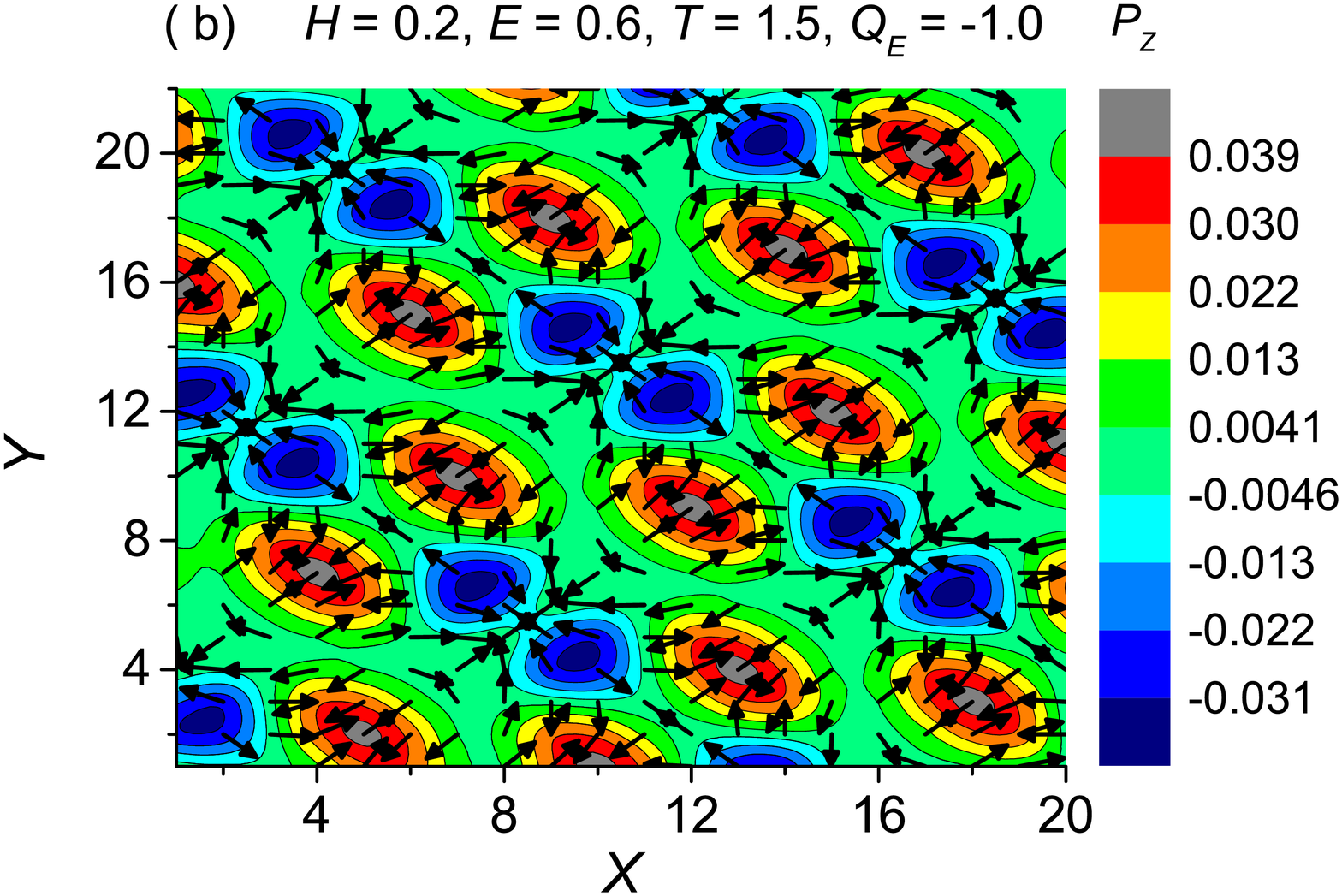,width=0.33\textwidth,height=5.2cm,clip=}
 \epsfig{file=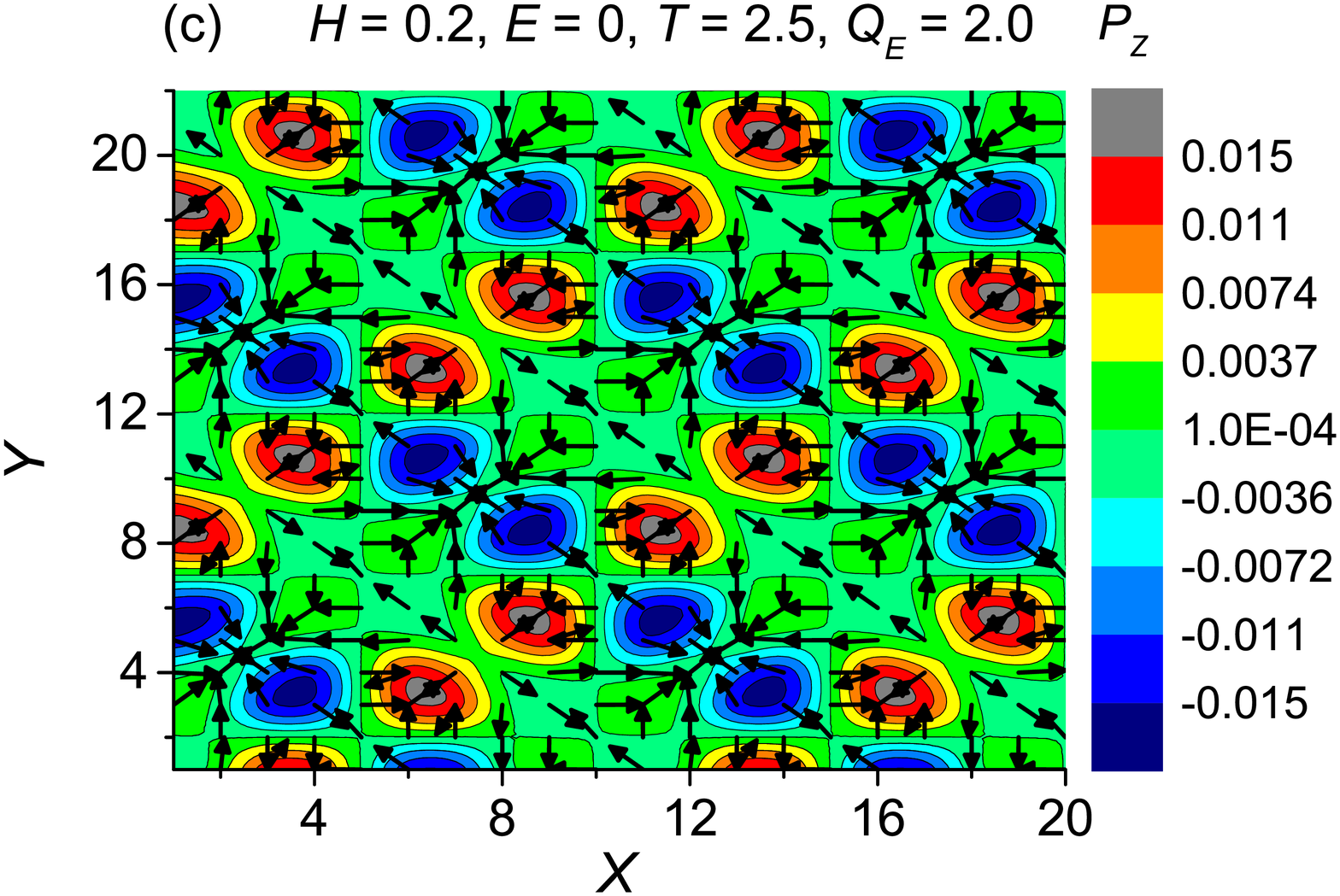,width=0.33\textwidth,height=5.2cm,clip=}
 }
 \vspace{-0.3cm}
\begin{center}
\parbox{16.5cm}{\small{{\bf Figure 4.}  FE  SL calculated at  (a) $H$ = 0.2, $E$ = 0, $T$ = 2.3,  (b) $H$ = 0.2, $E$ = 0.6, $T$ = 1.5,  and (c) $H$ = 0.2, $E$ = 0, $T$ = 2.5, respectively. }}
\end{center}
\end{figure*}
   At   first glance, the FE SLs depicted in  Figure 4(a) and (c) look  very similar. However actually,   in each   FE skyrmion of Figure 4(a),   the $xy$-components  in one blue area  are mirror-symmetric   to  another  blue area, but  those in the two red regions show  no symmetry;  whereas in  Figure 4(a) and (c),  the $xy$-components in the blue and red areas of every FE skymion are all mirror-symmetric with   respect to another corresponding area.

\subsection{Topological Charge Density Distributions of  Ferroelectric Skyrmionic Crystals}

 \vspace{-0.3cm}
The   $\rho_E$ values  of  every  FE SL also form  periodic crystal  as shown in Figure 5.  Since  each  FE skyrmion is stabilized by an  FM skyrmion, and  the numbers of the two sorts of skyrmions observed   in a square lattice chosen for simulations are  same, it is easy to infer  that:  in  the (a) case,  a unit cell of  the $\rho_E$ crystal contains one red,  one yellow, two pink and  three blue areas; in the (b) case,  the unit cell   has  one red,   two yellow and  five blue areas; and the last one includes  one red, one green and two yellow areas, respectively. These grouped  round colored areas appear periodically,   though   surrounded by much vaster colored background,  and the  $\rho_E$ crystals  look very distinct from each other.  The FE SLs displayed  in Figures 4(a) and (c) look very similar. But we see  here that the $\rho_E$ values  of most  sites     are   positive in Figures 5(c),   so  that its averaged topological charge $Q_E$ takes the  largest value.
\begin{figure*}[htb]
\centerline{
\epsfig{file=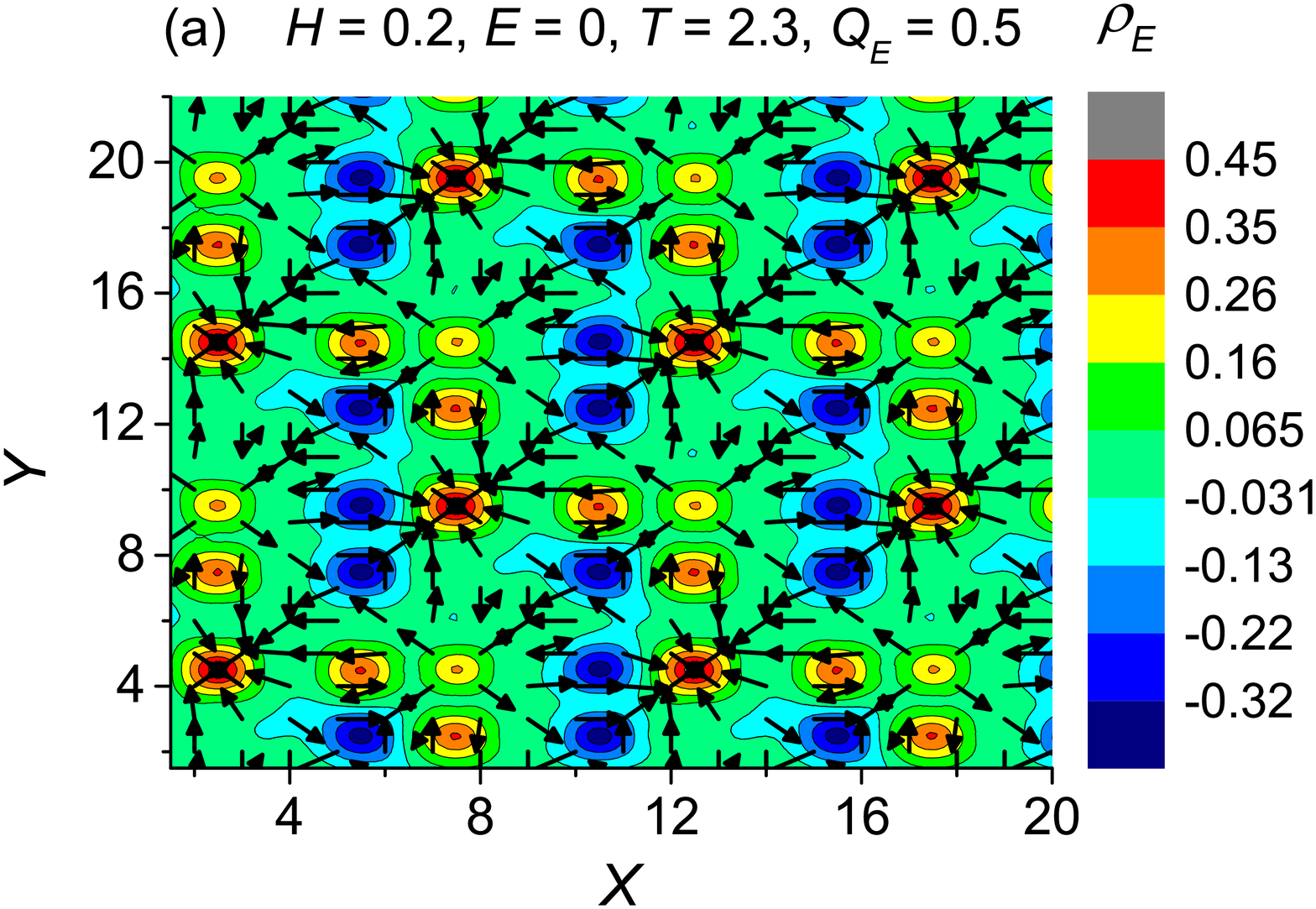,width=0.33\textwidth,height=5.2cm,clip=}
 \epsfig{file=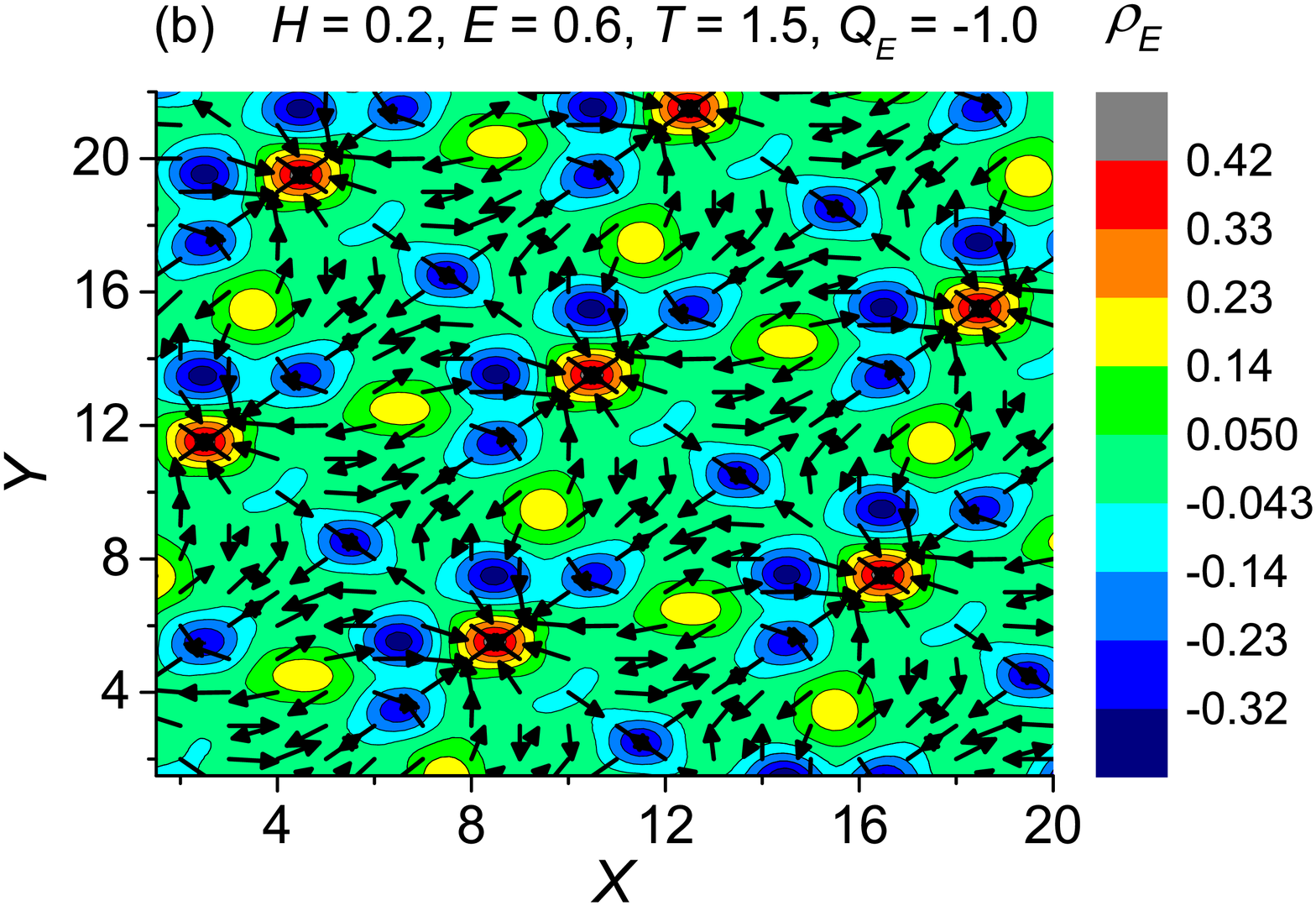,width=0.33\textwidth,height=5.2cm,clip=}
 \epsfig{file=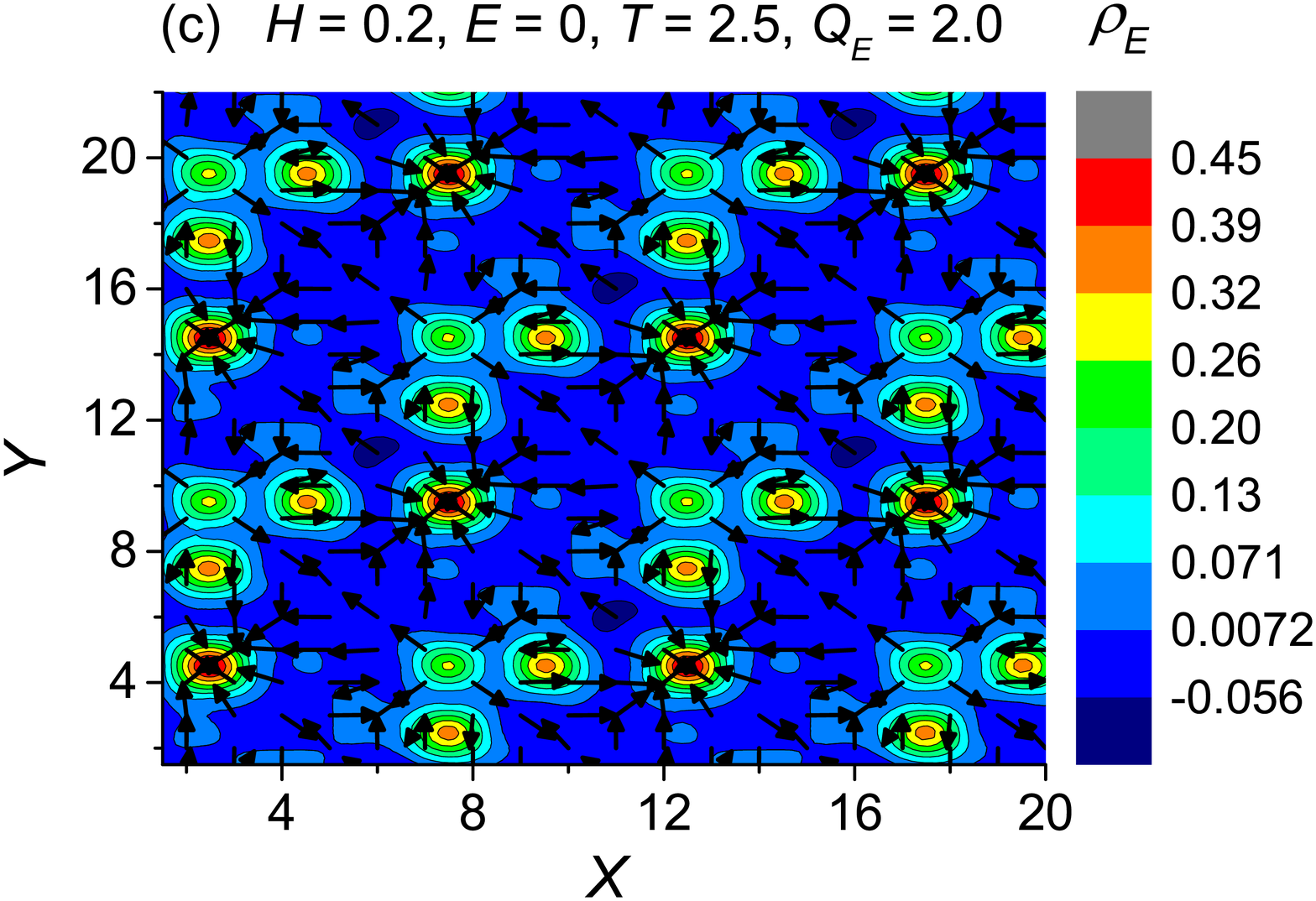,width=0.33\textwidth,height=5.2cm,clip=}
 }
\begin{center}
\vspace{-0.3cm}
\parbox{16.5cm}{\small{{\bf Figure 5.} The $\rho_E$ crystals  calculated  at   (a) $H$ = 0.2, $E$ = 0, $T$ = 2.3,  (b) $H$ = 0.2, $E$ = 0.6, $T$ = 1.5,  and (c) $H$ = 0.2, $E$ = 0, $T$ = 2.5, respectively.}}
\end{center}
\end{figure*}

\vspace{-0.5cm}
\subsection{Three-Dimensional  Diagrams  of  Individual Ferroelectric Skyrmions}

\vspace{-0.2cm}
 Figure 6 displays the  three-dimensional (3D)  diagrams of  the FE
\begin{figure*}[htb]
\centerline{
\epsfig{file=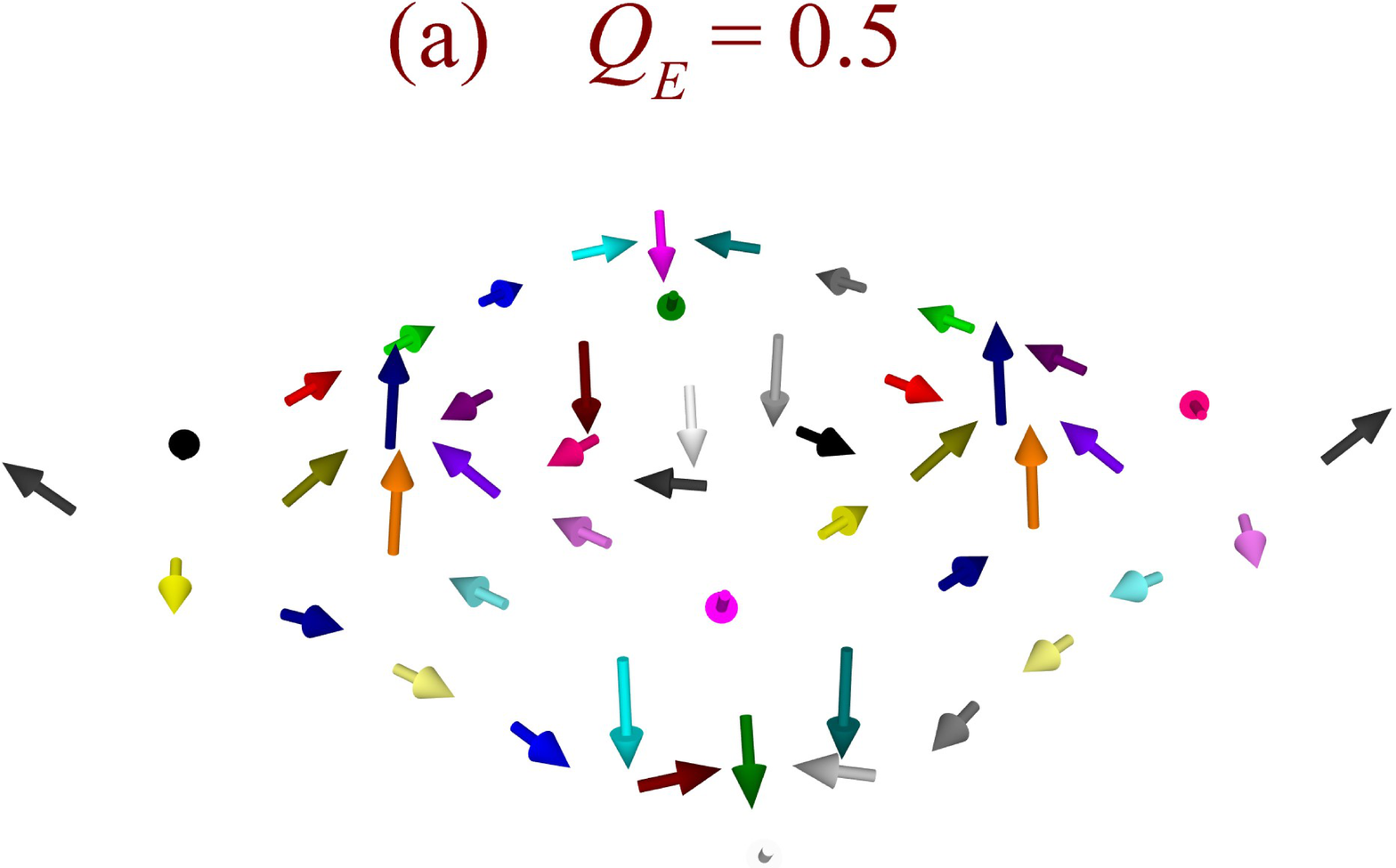,width=0.33\textwidth,height=3.5cm,clip=}
\epsfig{file=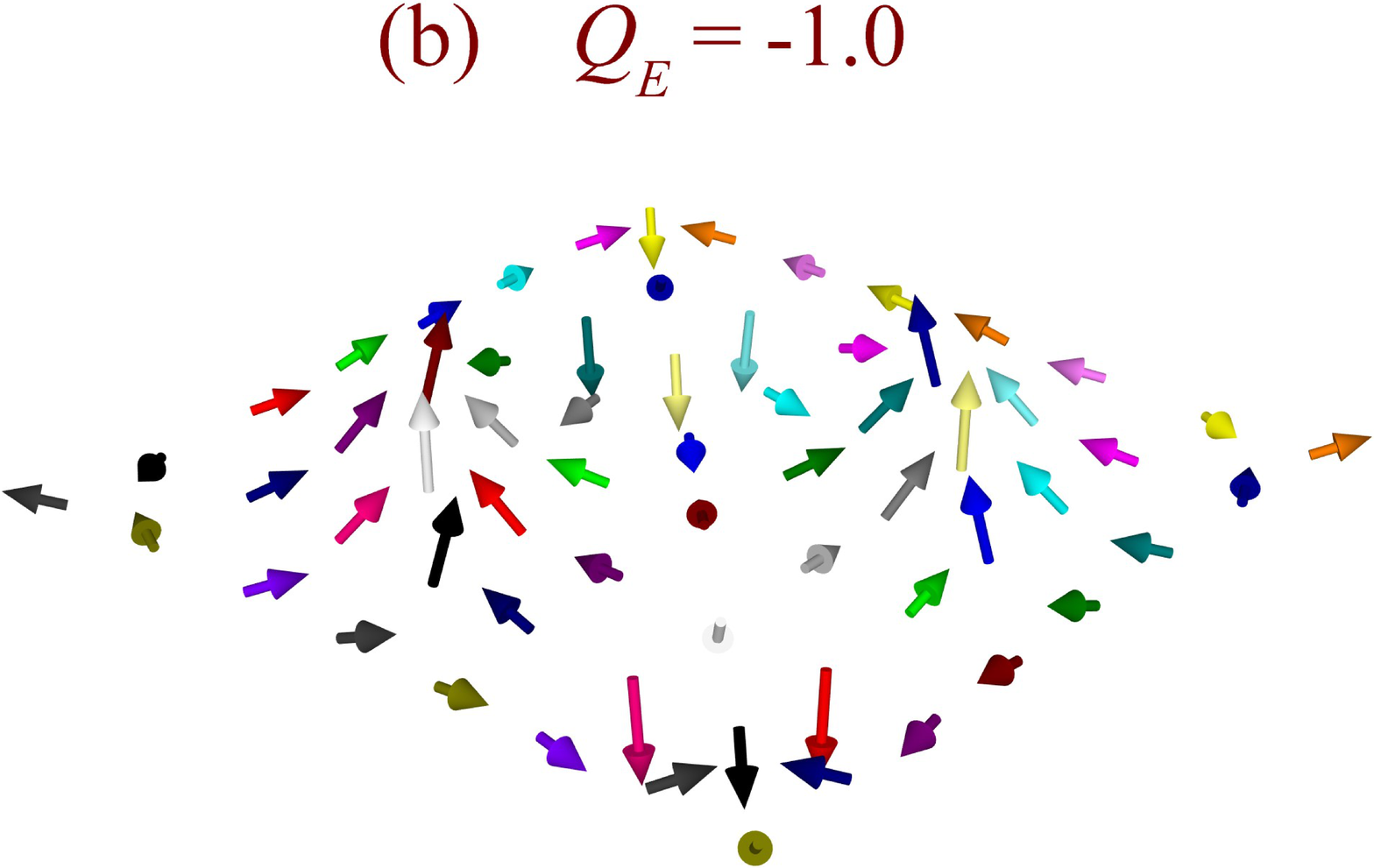,width=0.33\textwidth,height=3.5cm,clip=}
\epsfig{file=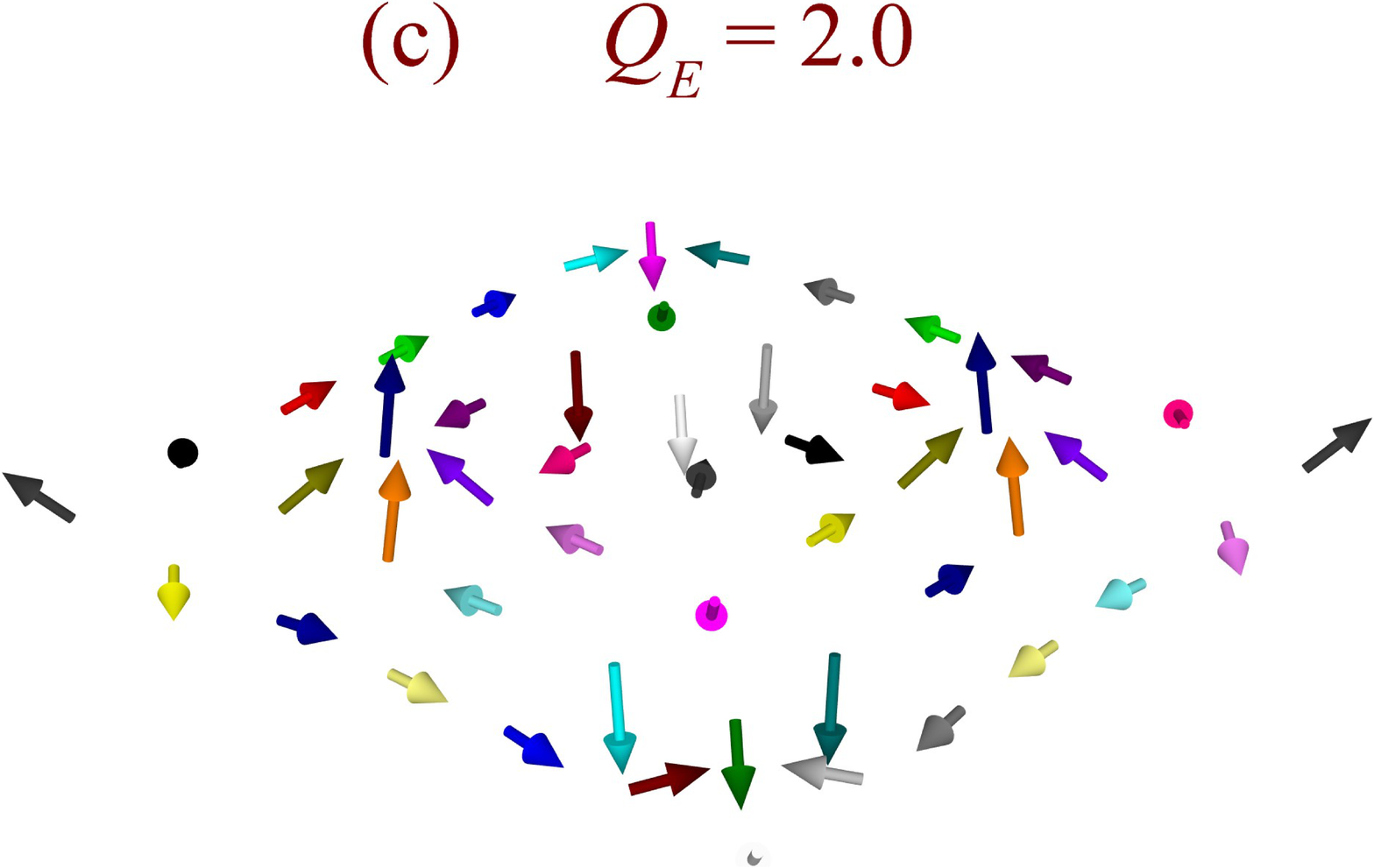,width=0.33\textwidth,height=3.5cm,clip=}
 }
 \vspace{-0.3cm}
\begin{center}
\parbox{16.5cm}{\small{{\bf Figure 6.}  Three-dimensional  individual  FE skyrmions calculated at  (a) $H$ = 0.2, $E$ = 0, $T$ = 2.3,  (b) $H$ = 0.2, $E$ = 0.6, $T$ = 1.5,  and (c) $H$ = 0.2, $E$ = 0, $T$ = 2.5, respectively.}}
\end{center}
\end{figure*}
skyrmions of    $Q_E$  = 0.5, -1.0 and 2.0 respectively.  Two peaks (up) and two hollowed (down) areas  observed in an FE skyrmion  correspond to the red and blue regions depicted  in  Figure 4.  Once again,   FE skyrmion  in  (b) obviously differ from other two, whereas the two depicted in (a) and (c) look quite similar.  However,     the $z$-components of the FE dipoles   in the whole   {\it 4}-th   column with the same $x$  coordinator are  actually  different  in signs, so that the topological charges   are  equal to  0.5 and 2.0  respectively   in the   two cases.

\section{ Conclusions and Discussion}
\vspace{-0.3cm}
In conclusion, we have calculated and characterized    the  topological textures of  FE SLs formed in a   2D   multiferroic  monolayer, where  DM interaction and ME coupling coexist,    by means of a quantum simulating approach.   In all cases,  the  topological charge per FM  skyrmion is  calculated  to be  -1.0. On the other hand, an FE SLs is formed  once  one FM SL is induced, and the topological charges  of the FE skyrmions in the   FE SLs are usually  quantized to be  integers, half integers and  multiples of certain fractional values. Each FE skymrion is  an electric dipole complex,  which is    stabilized by  one  FM skyrmion, so that the FE SL coincides with  the FM SL, and the    $\rho_E$ values  also form periodic pattern.

As already described above, an  normally applied electric fields is able to  change   skyrmion sizes,  and elevate the critical  temperature  $T_{SL}$, that is,  to  create FM and FE SLs above the original $T_{SL}$.  On the other hand, we have also found that the FM and FE SLs induced  by only  external magnetic fields    can be destroyed by  $E$-fields   applied afterwards. For example,  when $H$ = 0.2  and $E$ = 0,   FM and FE SLs are observed  below $T_{SL}$ = 3.15. If an $E$- field  of strength  0.6 is considered  afterwards and  the above calculated results are used as initial configurations to perform simulations,  as  $T \in$ (3.05, 2.5),   the  FM and FE SLs previously produced  disappear,  being replaced by  FM and FE helical crystals;   below that  $T$-range, the   FM  and FE SLs  are able to  maintain their   main structures as before, but the FM and FE skyrmions in the crystals are found to be  elongated by the $E$-field;  it is only at $T$ = 3.1, which  is immediately below the  original $T_{SL}$,  that FM  and FE SLs of    different crystal pattern can be stabilized.  In comparison,  when both the magnetic and electric fields of the above nonzero strengths are exerted simultaneously,   the  FM and FE SLs emerge if   $T \leq T_{SL}$ = 3.17  as shown in Figure 1(a). These  two sets of   results are completely different. However,   this phenomenon is  not strange,  since  the  properties of magnetic systems strongly depend on the magnetizing  processes.

The topological chargers of magnetic skyrmions and bimerons are all nonzero integers, and that of a magnetic vortex is equal to 0.5. However here,   the  FE skyrmions can be characterized  by topological  numbers of no only  integers, half integers, but also the  multiples of certain fractional values. This finding is  interesting, and may be of importance,  the physics behind needs to be further probed in depth.

\vspace{0.5cm}
 \centerline{\bf Acknowledgements}
\vspace{0.2cm}
The author  thanks the financial support provided by National Natural Science Foundation of China  under grant No.~11274177.


\begin{thebibliography}{35}

\bibitem{Bogdanov89} A.~N. Bogdanov, and D.~A. Yablonskii, Sov.~Phys. JETP, 95 (1989)

\bibitem{Bogdanov94} A.   Bogdanov,   and   A. Huber,
{J. Magn. Magn. Mater.} 138, 255-269 (1994)

\bibitem{Binz} B.~Binz, A.~ Vishwanath, V.~Aji, Phys. Rev.~Lett. 96,  207202 (2006).


\bibitem{Muhlbauer} S.~Muhlbauer, B.~Binz, F. Jonietz, C. Plfleiderer, A. Rosch, A. Neubauer, R.~Georgii, P. Boni, Science 323, 5916 (2009).

\bibitem{4Nagaosa} N. Nagaosa, and Y. Tokura, Nat. Nanotechnol. 8, 899 (2013).

\bibitem{12Fert13} A. Fert, V. Cros, and J. Sampaio, Nat. Nanotechnol. 8, 152
(2013).

\bibitem{13Dzyaloshinskii} I. E. Dzyaloshinskii, Sov. Phys. JETP 5, 1259 (1957).

\bibitem{14Moriya}T. Moriya, Phys. Rev. 120, 91 (1960).

\bibitem{15Fert80} A. Fert, and P. M. Levy, Phys. Rev. Lett. 44, 1538 (1980).

\bibitem{16Fert90} A. Fert, Metallic Multilayers 59-60, 439 (1990).

\bibitem{17Crepieux} A. Cr\'epieux,  and C. Lacroix, J. Magn. Magn. Mater. 182, 341 (1998).

\bibitem{Pappas}   C. Pappas,  E.  Lelievre-Berna,  P.  Falus,  P. M.
Bentley,  E.  Moskvin, S.    Grigoriev,  P. Fouquet,    and  B.
  Farago,  
  {
Phys. Rev. Lett.} 102, 197202 (2009).

\bibitem{Yu10}  X.~Z.    Yu, Y.   Onose,   N.  Kanazawa,  J. H.  Park,  J. H.  Han, Y.  Matsui,  N.  Nagaosa,    and  Y.  Tokura,
    {  Nature} 465, 901 (2010).

\bibitem{Banerjee}  S.     Banerjee, J. Rowland,   O.    Erten,  and   M.
Randeria,
 { Phys. Rev. X} 4, 031045 (2014).

\bibitem{Yi}   S.~D. Yi,   S.  Onoda, N.   Nagaosa,   and  J. H. Han,
{ Phys. Rev. B} 80, 054416 (2009).

\bibitem{Buhrandt} S.   Buhrandt,   and   L.  Fritz,
{Phys. Rev. B }88, 195137 (2013).

\bibitem{Huang}  S. X.  Huang,  and  C. L.  Chien,
Phys.~Lett. 108, 267201 (2012).

\bibitem{Romming15} N.    Romming,   A.  Kubetzka, C.  Hanneken,  K.   von
Bergmann,  and   R.   Wiesendanger,
{ Phys. Rev. Lett.} 114, 177203 (2015).

\bibitem{Iwasaki} J.   Iwasaki,   M.  Mochizuki,  and   N.    Nagaosa,
 { Nat. Commun.}  {  4},  1463  (2013).

\bibitem{Jonietz} F. Jonietz, S. M\"uhlbauer, C. Pflerderer, A. Neubauer, W. M\"unzer, A. Bauer, T. Adams, R. Georgii, P. B\"oni, R. A. Duine, K. Everschor, M. Garst, and A. Rosch, Sience 330, 1648 (2010).

 \bibitem{Schultz} T. ~Schultz, R. Ritz, A. Bauer, M. Halder, M. Wagner, C. Franz, C. Pflerderer, K. Everschor, M. Garst, and A. Rosch,
     Nat. Phys. 8, 301 (2012).

\bibitem{Everschor} K. Everschor, M. Garst,  R. A. Duine, and A. Rosch, Phys. Rev. B 84, 064401  (2011).

\bibitem{Zang} J. Zang, M Mostovoy, J. H. Han, and N. Nagaosa, Phys. Rev. Lett. 107, 136804  (2011).

\bibitem{Seki12} S. Seki, S.~ Ishiwata, and Y.~Tokura, Phys. Rev. B 86, 060403 (R) (2012).

 \bibitem{LiuYH} Y.~H. Liu, Y.~Q. Li, and J. ~H. Han, Phys. Rev. B 87, 100404 (2013).

 \bibitem{LiuYH2} Y.~H. Liu,  J. H. Han,  A.~A. Omrani, H.~M.~Ronnow,and Y.~Q. Li,  
     arXiv:1310.5293v1 [cond-mat.str-el] 20 Oct 2013


 \bibitem{JiaCL} C. Jia, S.~Onoda, N. Nagaosa, and J.H. Han, Phys. Rev. B 76, 144424  (2007).

\bibitem{Tokura}  Y. Tokura, S. Seki, and Y. Tokura, arxiv: 1206.4404v1 (2012).


 \bibitem{LiuIan20-SM} Z .-S. Liu, H. Ian,
 { Superlattices Microstruct.} {  138}, 106379  (2020) .



\bibitem{liujpcm} Z.-S.    Liu,  V.    Sechovsk{\' y},   and   M.    Divi{\v
s}, 
 { J. Phys.: Condens. Matter}  23, 016002 (2011).

\bibitem{liupssb}  Z.-S.    Liu,  V.    Sechovsk{\' y},   and   M.    Divi{\v s}, 
    { Phys. Status Solidi B}  249, 202  (2012).




\bibitem{LiuIan19-2} Z.-S.  Liu,   and  H.    Ian,
{ J. Phys.: Condens. Matter.} {  31}, 29 (2019).

\bibitem{wikipedia} Magnetic skyrmion,  https://en.wikipedia.org/wiki/Magnetic\_skyrmion

\bibitem{Berg} B. Berg, M. L\"uscher, Nuclear Phys. B 190[FS3] , 412 (1981)

 \end{thebibliography}
 \end{document}